\documentclass[iop]{emulateapj}
\usepackage{apjfonts} 
\usepackage{graphicx}
\def\alwaysmath#1{\ifmmode{#1}\else{$#1$}\fi} 
\usepackage{rotating}

\usepackage[pdfpagemode={UseOutlines},bookmarks=true,bookmarksopen=true,
   bookmarksopenlevel=0,bookmarksnumbered=true,hypertexnames=false,
   colorlinks,linkcolor={blue},citecolor={blue},urlcolor={red},
   pdfstartview={FitV},unicode,breaklinks=true]{hyperref}
 
\lefthead{Hasselquist et al.} 
\righthead{APOGEE [C/N] Abundances Across the Galaxy}
\begin{document} 
\DeclareGraphicsExtensions{.ps,.pdf,.png,.jpg,.eps}

\title{APOGEE [C/N] Abundances Across the Galaxy: Migration and Infall from Red Giant Ages} 
 
\author{Sten Hasselquist\altaffilmark{1}, 
Jon A. Holtzman\altaffilmark{1}, 
Matthew Shetrone\altaffilmark{2}, 
Jamie Tayar\altaffilmark{3}, 
David H. Weinberg\altaffilmark{3}, 
Diane Feuillet\altaffilmark{5},
Katia Cunha\altaffilmark{6}, 
Marc H. Pinsonneault\altaffilmark{3}, 
Jennifer A. Johnson\altaffilmark{3}, 
Jonathan Bird\altaffilmark{4}, 
Timothy C. Beers\altaffilmark{7},
Ricardo Schiavon\altaffilmark{14}, 
Ivan Minchev\altaffilmark{13},
J. G. Fern\'andez-Trincado\altaffilmark{8,9,10},
D. A. Garc{\'{\i}}a-Hern{\'a}ndez\altaffilmark{11,12},
Christian Nitschelm\altaffilmark{15},
Olga Zamora\altaffilmark{11,12}
}

% Verne Smith\altaffilmark{3}, 
% Andrew McWilliam\altaffilmark{4}, 
% J. G. Fern\'andez-Trincado\altaffilmark{8},
% Timothy C. Beers\altaffilmark{7},
% Steven R. Majewski\altaffilmark{5},   
% David L. Nidever\altaffilmark{3}, 
% Baitian Tang\altaffilmark{8},
% Patricia B. Tissera\altaffilmark{19},
% Emma Fern{\'a}ndez Alvar \altaffilmark{17},
% Carlos Allende Prieto\altaffilmark{9,10},
% Giuseppina Battaglia\altaffilmark{9,10},
% Leticia Carigi\altaffilmark{17},
% Katia Cunha\altaffilmark{12}, 
% Gloria Delgado-Inglada \altaffilmark{17},
% Peter Frinchaboy\altaffilmark{6}, 
% D. A. Garc{\'{\i}}a-Hern{\'a}ndez\altaffilmark{9,10},
% Doug Geisler \altaffilmark{16},
% Dante Minniti\altaffilmark{13,14,15},
% Vinicius M. Placco\altaffilmark{7}, 
% Mathias Schultheis\altaffilmark{18},
% Jennifer Sobeck\altaffilmark{5},
% Sandro Villanova\altaffilmark{8}

\altaffiltext{1}{New Mexico State University, Las Cruces, NM 88003, USA (sten@nmsu.edu, holtz@nmsu.edu)}

\altaffiltext{2}{University of Texas at Austin, McDonald Observatory, Fort Davis, TX 79734, USA (shetrone@utexas.edu)}

\altaffiltext{3}{Department of Astronomy, The Ohio State University, Columbus, OH 43210, USA (tayar@astronomy.ohio-state.edu, pinsonneault.1@osu.edu, johnson.3064@osu.edu, dhw@astronomy.ohio-state.edu)}

\altaffiltext{4}{Department of Physics and Astronomy, Vanderbilt University, Nashville, TN 37235, USA (jonathan.bird@vanderbilt.edu)}

\altaffiltext{5}{Max-Planck-Institut f{\"u}r Astronomie, K{\"o}nigstuhl 17, D-69117 Heidelberg, Germany (feuilldk@gmail.com)}

\altaffiltext{6}{Observat\'orio Nacional/MCTI, Rua Gen. Jos\'e Cristino, 77, 20921-400, Rio de Janeiro, Brazil (cunha@email.noao.edu)}

\altaffiltext{7}{Department of Physics and JINA Center for the Evolution of the Elements, University of Notre Dame, Notre Dame, IN 46556, USA  (tbeers@nd.edu)}

\altaffiltext{8}{Departamento de Astronom\'ia, Casilla 160-C, Universidad de Concepci\'on, Concepci\'on, Chile  (jfernandezt@astro-udec.cl)}

\altaffiltext{9}{Instituto de Astronom\'ia y Ciencias Planetarias, Universidad de Atacama, Copayapu 485, Copiap\'o, Chile.}

\altaffiltext{10}{Institut Utinam, CNRS UMR 6213, Universit\'e Bourgogne-Franche-Comt\'e, OSU THETA Franche-Comt\'e, Observatoire de Besan\c{c}on, \\ BP 1615, 25010 Besan\c{c}on Cedex, France.}

\altaffiltext{11}{Instituto de Astrof{\'{\i}}sica de Canarias, E-38205 La Laguna, Tenerife, Spain (agarcia@iac.es)}

\altaffiltext{12}{Departamento de Astrof{\'{\i}}sica, Universidad de La Laguna (ULL), E-38206 La Laguna, Tenerife, Spain (agarcia@iac.es)}

\altaffiltext{13}{Leibniz-Institut f{\"u}r Astrophysik Potsdam (AIP), An der Sternwarte 16, D-14482, Potsdam, Germany (iminchev1@gmail.com)}

\altaffiltext{14}{Astrophysics Research Institute, Liverpool John Moores University, 146 Brownlow Hill, Liverpool L3 5RF, UK (rpschiavon@gmail.com)}

\altaffiltext{15}{Centro de Astronom{\'{\i}}a (CITEVA), Universidad de Antofagasta, Avenida Angamos 601, Antofagasta 1270300, Chile (christian.nitschelm@uantof.cl)}

\begin{abstract} 
We present [C/N]-[Fe/H] abundance trends from the SDSS-IV Apache Point Observatory Galactic Evolution Experiment (APOGEE) survey, Data Release 14 (DR14), for red giant branch stars across the Milky Way Galaxy (MW, 3 kpc $<$ R $<$ 15 kpc). The carbon-to-nitrogen ratio (often expressed as [C/N]) can indicate the mass of a red giant star, from which an age can be inferred. Using masses and ages derived by Martig et al., we demonstrate that we are able to interpret the DR14 [C/N]-[Fe/H] abundance distributions as trends in age-[Fe/H] space. Our results show that an anti-correlation between age and metallicity, which is predicted by simple chemical evolution models, is not present at any Galactic zone. Stars far from the plane ($|$Z$|$ $>$ 1 kpc) exhibit a radial gradient in [C/N] ($\sim$ $-$0.04 dex/kpc). The [C/N] dispersion increases toward the plane ($\sigma_{[C/N]}$ = 0.13 at $|$Z$|$ $>$ 1 kpc to $\sigma_{[C/N]}$ = 0.18 dex at $|$Z$|$ $<$ 0.5 kpc). We measure a disk metallicity gradient for the youngest stars (age $<$ 2.5 Gyr) of $-$0.060 dex/kpc from 6 kpc to 12 kpc, which is in agreement with the gradient found using young CoRoGEE stars by Anders et al. Older stars exhibit a flatter gradient ($-$0.016 dex/kpc), which is predicted by simulations in which stars migrate from their birth radii. We also find that radial migration is a plausible explanation for the observed upturn of the [C/N]-[Fe/H] abundance trends in the outer Galaxy, where the metal-rich stars are relatively enhanced in [C/N]. 
\end{abstract}

 
\section{Introduction}

The SDSS-IV Apache Point Observatory Galactic Evolution Experiment (APOGEE, \citealt{Majewski2017}) survey provides detailed chemical abundances of 18 chemical elements for $\sim$ $10^{5}$ red giant stars across the disk of the Milky Way (MW) Galaxy (3-15 kpc). From this extensive data set, maps of the chemical abundance distributions of the disk can be generated. Valuable information about the chemical evolution and star-formation history of the MW disk can be inferred from these maps, as done in the APOGEE ``chemical cartography'' program (see  \citealt{Hayden2014,Hayden2015}; J.\ Holtzman et al., in prep.). However, unlike most chemical abundances provided by APOGEE, the abundances of carbon and nitrogen in red giant stars are not necessarily a reflection of the stellar primordial abundance as the observed ratio of these elements can be altered due to stellar evolution, thus complicating the interpretation of C and N abundance maps across the Galaxy. 

It has been shown that the atmospheric C-to-N abundance ratio that a star is formed with (often expressed as [C/N]) changes as a star evolves from the main sequence to the red giant branch (see, e.g., \citealt{Iben1965,Salaris2015}). As the star ascends the giant branch, the convective layer extends towards the core where CN-cycle processed material lies. This material is enhanced in N relative to the star's outer layers. The depth of this layer, and therefore how much nitrogen-enhanced material is dredged up, depends on the mass of the star. Consequently, the observed [C/N] abundance for a star on the giant branch can indicate the \emph{mass} of the star, from which an \emph{age} can be inferred by invoking models for stellar evolution. 

This phenomenon has been exploited in the APOGEE sample to show that the $\alpha$-element enhanced stars in the MW disk are generally older than the disk stars with Solar $\alpha$-element abundance \citep{Masseron&Gilmore2015}, although there exist a few $\alpha$-element enhanced young stars in the Solar Neighborhood (see, e.g., \citealt{Chiappini2015} and \citealt{Martig2015}). Additionally, \citet{Martig2016a} and \citet{Ness2016} have derived ages for the APOGEE DR12 sample using masses from asteroseismology, showing that there is an age gradient in the geometrically-defined thick disk \citep{Martig2016b}, indicating that this structure is not uniformly old across all Galactocentric radii. 

The prospect of [C/N] abundance as an age indicator opens the possibility for studying [C/N]-metallicity or \emph{age}-metallicity abundance trends across the MW disk. Simple, closed-box and leaky-box chemical evolution models suggest that the Galactic interstellar medium (ISM) should become more enriched in metals over time as stars/supernovae synthesize metals and expel some fraction of them back to the ISM. This implies that stars born at more recent times from this enriched material should be more metal-rich than stars born at older times. These simple models predict that there should be a correlation between age and metallicity for stars in the MW. 

Initially, an age-metallicity relation in the Solar Neighborhood was found through Str{\"o}mgren photometry of F stars (see, e.g., \citealt{Twarog1980,Carlberg1985}). However, a multitude of chemical abundance studies of sub-giants in the Solar Neighborhood have shown that there is a lack of an age-metallicity relation, with large scatter in age at any given metalliticy (see, e.g., \citealt{Edvardsson1993,Bensby2004,Haywood2013,Feuillet2016}). Simulations suggest that an age-metallicity relation can be ``smeared'' out by radial mixing (see, e.g., \citealt{Minchev2013}). Such mixing could be responsible for the metal-rich stars in the Solar Neighborhood that may be older than the more metal-poor stars (see, e.g., \citealt{Fuhrmann2008}). In fact, it may be the case that large fractions of stars in the Solar Neighborhood actually migrated in from a different birth location \citep{Chiappini2014}. \citet{Feuillet2018} find that the mean age of Solar Neighborhood stars is youngest near solar metallicity and increases for both sub-solar and super-solar populations. 

Radial migration, which changes a star's orbital radius without strongly affecting its eccentricity, could occur through a variety of dynamical mechanisms.  Most notably \citet{Sellwood2002} showed that spiral arm transients could cause a star to change its guiding center radius while remaining on a near circular orbit, and they argued that this radial-mixing mechanism should operate generically in disk galaxies. In their chemical evolution models with radial mixing, \citet{Schonrich&Binney2009} distinguish between ``blurring,'' overlap of stars in radial zones because of orbital eccentricity, and ``churning'' that changes guiding center radii. \emph{N}-body simulations show that churning is an important phenomenon in dynamically realistic disks, though the degree of mixing depends on details such as gas richness, bar formation, and satellite perturbations (e.g., \citealt{Roskar2008,Bird2013,Minchev2013}).  It has also been suggested that radial migration could play a key role in thick-disk formation (e.g., \citealt{Schonrich&Binney2009b,Loebman2011}). However, \citet{Minchev2013} find it more likely that heating by mergers is responsible for the observed properties of the inner thick disk, which consists of stars born hot in a turbulent gas-rich phase at early times (see also \citealt{Minchev2017,Minchev2018}). 

Two major observational constraints on the extent to which radial migration affected the spatial distributions of the MW stellar populations are the radial variation of the observed metallicity distribution functions (MDFs) and the radial variation of age-metallicity relations. \citet{Hayden2015} invoked a simple radial-migration model to explain the observed MDFs of stars observed by APOGEE across the MW. They found that such a model could explain the negatively-skewed MDF in the inner Galaxy and the positively-skewed MDF for the outer Galaxy. \citet{Loebman2016} were able to match the change in skewness of the APOGEE MDFs with their high-resolution, \emph{N}-body+smooth particle hydrodynamics (SPH) simulation by including a prescription for radial migration.

While attempts have been made to match the observed age-metallicity relation (or lack thereof) of the Solar Neighborhood using simulations with radial migration, there are few observational constraints on the age-metallicity relations across other regions of the Galaxy. Some simulations predict that there should be an increase in the relative number of outward migrators with Galactic radius, which leads to a flattening of the expected age-metallicity anti-correlation with increasing Galactic radius (e.g., \citealt{Minchev2014}). As large spectroscopic surveys continue to provide precise chemical abundances for stars across the Galaxy, attempts can be made to quantify radial migration in the MW, which has important implications for understanding galaxy evolution and the observed lack of age-metallicity relations (e.g., \citealt{Anders2018,Minchev2018}).   

In this paper we present [C/N]-[Fe/H] abundance results from the latest APOGEE data release (DR14) for stars distributed across the MW disk (3 kpc $<$ R $<$ 15 kpc). Similar to previous work, we use the [C/N] abundance of a red giant star as an indicator of its age. This work marks the first time [C/N]-[Fe/H] abundance trends of the MW have been presented in this way (individual stars rather than mean abundance/age maps), allowing for an exploration of age-metallicity relations across much of the Galaxy. From these spatially resolved age-metallicity relations, we are able to characterize the near-present day metallicity of the ISM across the Galaxy, as well as analyze how radial migration played a role in shaping these age-metallicity relations. We present and interpret [C/N]-[Fe/H] abundance trends rather than converting to age-metallicity space to avoid introducing systematic uncertainties through a [C/N]-age fit. This also allows us to explore other potential interpretations of [C/N] abundance variations across the Galaxy, which would be masked in a study that converts [C/N] to age.

%Similar to previous work, we use [C/N] as an indicator for the ages of red giant stars to interpret these [C/N]-[Fe/H] abundance trends as age-[Fe/H] trends. 

%However, rather than use inferred ages, which have both statistical and systematic uncertainties, we examine trends directly in abundance space, where the observational errors are relatively small. We analyze how these abundance tracks change as a function of height above the Galactic plane and distance from the Galactic center.  

The observations are described in \S \ref{sec:obs}. Sample selection and relating the DR14 [C/N] abundance values to mass/age are discussed in \S \ref{sec:sample}. We present the [C/N] abundance results in \S \ref{sec:res} and interpret them in the context of stellar age and Galactic evolution in \S \ref{sec:disc}.

\section{Observations, Data Reduction, and Analysis}
\label{sec:obs}

The APOGEE survey was part of Sloan Digital Sky Survey III \citep{Eisenstein2011}, and observed 146,000 stars in the Milky Way galaxy \citep{Majewski2017} from 2011-2014. APOGEE-2 began observations in 2014 as part of the Sloan Digital Sky Survey IV \citep{Blanton2017} and DR14 contains an additional $\sim$ 100,000 stars \citep{dr142018}. The APOGEE instrument is a high-resolution (R $\sim$ 22,500) near-infrared (1.51-1.70 $\mu$m) spectrograph described in detail in Wilson et al. (in prep). The instrument was connected to the Sloan 2.5m telescope \citep{Gunn2006} and targets were observed according to \citet{Zasowski2013} for APOGEE, and \citet{Zasowski2017} for APOGEE-2.

The APOGEE data are reduced through methods described by \citet{Nidever2015}, and stellar parameters/chemical abundances are extracted using the APOGEE Stellar Parameters and Chemical Abundances Pipeline (ASPCAP, \citealt{Garcia2016}). ASPCAP interpolates in a grid of synthetic spectra \citep{Zamora2015} to find the best fit (through $\chi^{2}$ minimization) to the observed spectrum by varying T\raisebox{-.4ex}{\scriptsize eff}, surface gravity, microturbulence, metallicity, carbon abundance, nitrogen abundance, and $\alpha$-element abundance. In this analysis we use results from the 14th data release of SDSS (DR14, \citealt{dr142018} and \citealt{Holtzman2018}). Chemical abundances are derived from lines described in \citet{Shetrone2015}.  The C and N abundances are derived from features of CO and CN molecules, the data for which come from \citet{Pickering1996} and \citet{Sneden2014}, respectively. 

In this work, we only analyze APOGEE MW stars that satisfy the following quality cuts to further ensure reliable parameter/abundance derivation:
\begin{itemize}
  \item{S/N $>$ 80}
  \item{3500 K $<$ T\raisebox{-.4ex}{\scriptsize eff} $<$ 5500 K}
  \item{Log(g) $>$ 1.0}
  \item{No ASPCAPBAD flag set\footnote{This flag is set if ASPCAP finds an error in deriving any stellar parameters. It is fully described on the DR14 webpages (http://www.sdss.org/dr14)}}
\end{itemize}

Uncertainties in the APOGEE chemical abundances are empirically derived by analyzing the abundance scatter in calibration clusters in bins of T\raisebox{-.4ex}{\scriptsize eff}, [M/H], and S/N. The typical precision of C and N abundances in the parameter space studied in this work is $\sim$ 0.03-0.05 dex. As described further in \S \ref{sec:sample}, we divide the APOGEE  sample into two groups along the giant branch. The warmer stars residing in the lower part of the giant branch have median C and N uncertainties of 0.04 and 0.06 dex, respectively, resulting in a median [C/N] uncertainty of 0.07 dex. The cooler stars residing in the upper part of the giant branch have smaller median C and N uncertainties of 0.02 and 0.03 dex, respectively, resulting in a median [C/N] uncertainty of 0.04 dex. Because of the ASPCAP grid edge, we do not have any stars in our sample with [N/Fe] $>$ +1.0 dex, which may result in the lowest [C/N] stars dropping out of our sample.

Masses and ages for APOGEE stars analyzed in this work come from \citet{Martig2016a}, who used the APOKASC sample \citep{Pinsonneault2014} to derive a relation between APOGEE DR12 stellar parameters (of which [M/H] and [C/N] are the primary drivers) and mass/age, which was then applied to the APOGEE DR12 red giant sample, delivering a catalog of $\sim$ 52,000 APOGEE stars with masses accurate to $\sim$ 14\% and ages accurate to $\sim$ 40\%. In this work, we adopt the \citet{Martig2016a} seismic masses and resultant ages for the $\sim$ 1,500 stars in the APOKASC sample to demonstrate that DR14 [C/N] traces mass as DR12 [C/N] did.  Distances are derived by methods described in \citet{Hayden2014}, and are accurate to $\sim$ 20\%. For the presentations of our abundance results at various radial zones of the Galaxy, we use standard rectangular Galactic coordinates (X,Y,Z) where the sun is placed at 8 kpc.

\section{[C/N] Analysis and Sample Selection}
\label{sec:sample}

To date, the APOGEE literature works that use [C/N] abundances as mass/age indicators for red giant stars (e.g.,  (\citealt{Masseron&Gilmore2015,Martig2016a,Ness2016}) have used data from the 12th SDSS data release (DR12, \citealt{Alam2015}). In this section, we first verify that the newer DR14 [C/N] abundances trace seismic mass in the same way as DR12 [C/N], and understand across what parameter space this relation is viable. Then we verify that DR14 [C/N] abundances trace stellar ages that are inferred from the seismic masses.

\subsection{DR14 [C/N] as a Mass Indicator}

The observed atmospheric [C/N] abundance of a red giant star is the primordial mixture modified by first dredge up, with potential additional mixing that further alters the [C/N] abundance as a star evolves past the red bump and/or undergoes the helium flash (see e.g, \citealt{Gratton2000,Masseron2017,Lagarde2018,Shetrone2018:inprep}). The extent to which the primordial [C/N] abundance varies across the Galaxy is not well-understood because it necessitates the study of dwarf and/or subgiant stars that have not yet undergone first dredge up. \citet{Martig2016a} showed that APOGEE subgiant stars, when separated by their location in the Galaxy, exhibited similar [C/N]-[Fe/H] abundance tracks, suggesting that the primordial [C/N] abundance variation is small (at least within a few kpc of the Sun). Independent studies of nearby solar twins find a small variation of $\sim$ 0.1 dex in [C/Fe]  (see, e.g., \citealt{Nissen2015}). This variation is thought to be due to some stars exhibiting deficiencies in refractory elements, potentially a result of planet formation (e.g., \citealt{Nissen2015,Spina2016,Nissen2017,Bedell2018}). If this is the reason for [C/Fe] variation, then the [C/N] abundance ratio should not vary, as C and N have similar condensation temperatures \citep{Lodders2003}. These studies suggest that the [C/N] variation we observe in the APOGEE red giant stars is likely dominated by mass variation across the Galaxy rather than primordial variation. We further explore how primordial variation might affect our interpretations in \S \ref{sec:AGB}. 
 
Additionally, the [C/N] abundance ratio of red giant stars may be altered after first dredge up via ``extra'' metallicity-dependent mixing that takes place as a star evolves along the giant branch past the red bump (see, e.g., \citealt{Gratton2000,Shetrone2018:inprep}). \citet{Lagarde2018} propose that thermohaline mixing can explain the changing [C/N] abundance as a star evolves further up the red giant branch. There is also some evidence that the [C/N] changes when a low-mass star moves in to the core-helium burning red clump (RC) phase after the helium flash \citep{Masseron2017}. Additionally, core-helium burning stars on the RC suffer from higher uncertainties in their asteroseismic masses, and may even suffer from [C/N] measurement systematics (see, e.g., \citealt{Tayar2017}). Because of these limitations on [C/N] as a mass indicator, we divide the APOGEE sample analyzed in this work into two groups along the red giant branch. The lower red giant branch (LGB) sample contains stars with 2.6 $<$ log(g) $<$ 3.3 and the upper red giant branch (UGB) sample contains stars with 1.0 $<$ log(g) $<$ 2.1. As explained in more detail below, we are more confident that [C/N] traces mass in the LGB sample, but we include the UGB sample because of its more extensive spatial coverage.

\begin{figure*}
	\epsscale{1.0}
	\plotone{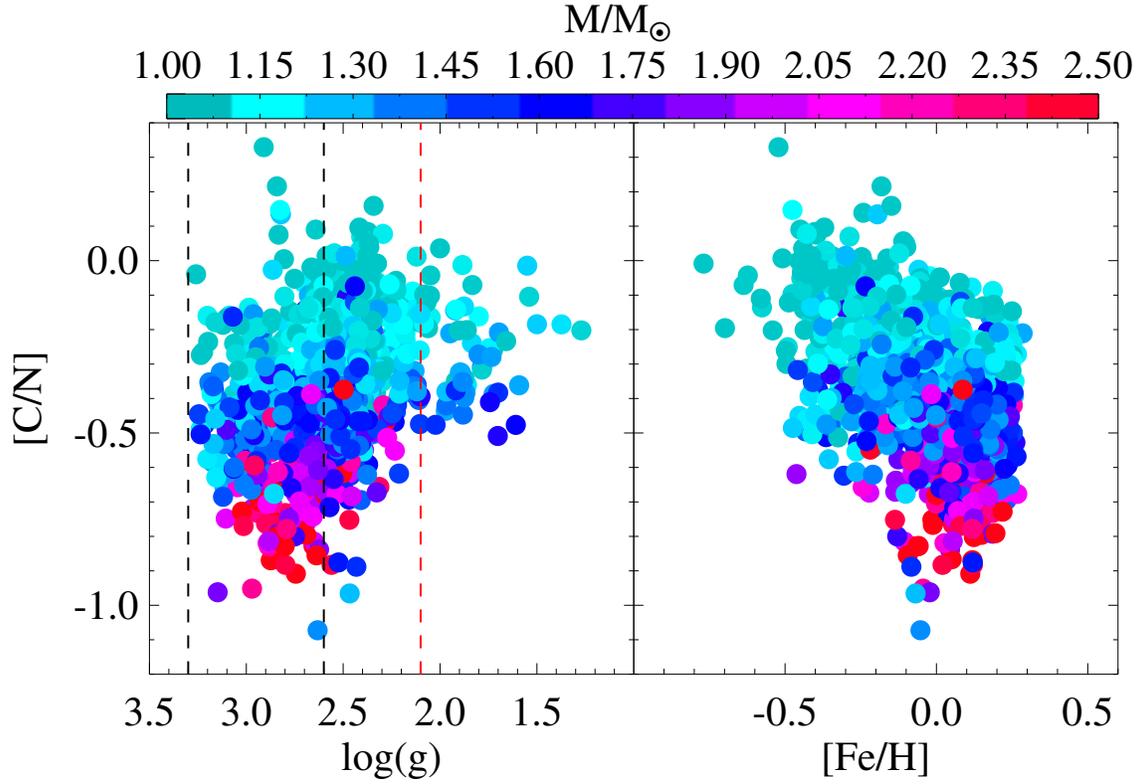} 
	\caption{[C/N] plotted as a function of log(g) (left) and [Fe/H] (right) for stars with seismic masses derived in \citet{Martig2016a}. The points are colored by seismic mass, and the black and red dashed lines on the left plots mark the divisions along the giant branch we use to study the abundance patterns of the MW.}
	\label{fig:martig_mass_plot}
\end{figure*}

The sample divisions are shown in Figure \ref{fig:martig_mass_plot}, where we plot DR14 [C/N] as a function of DR14 surface gravity and [Fe/H] for stars in the APOKASC sample. Points are colored by seismic mass provided in \citet{Martig2016a}. Qualitatively, we see that both LGB and UGB samples show a relation between [C/N] and mass, as seen by the color gradient. The lower cut in log(g) of the LGB sample serves to remove the RC stars, which may undergo extra mixing during the helium flash. However, we are sure to include secondary clump stars (2RC) in this sample. The 2RC stars are the most massive stars in our sample, and correspondingly have the lowest [C/N] values, as seen in the left panel of Figure \ref{fig:martig_mass_plot} (red points). Stars with M $\gtrsim$ 2.2 $M_{\odot}$ spend the majority of their evolved lifetime as 2RC stars rather than RGB stars, which is likely why there are no massive stars in the UGB sample.

\begin{figure*}
	\epsscale{1.0}
	\plotone{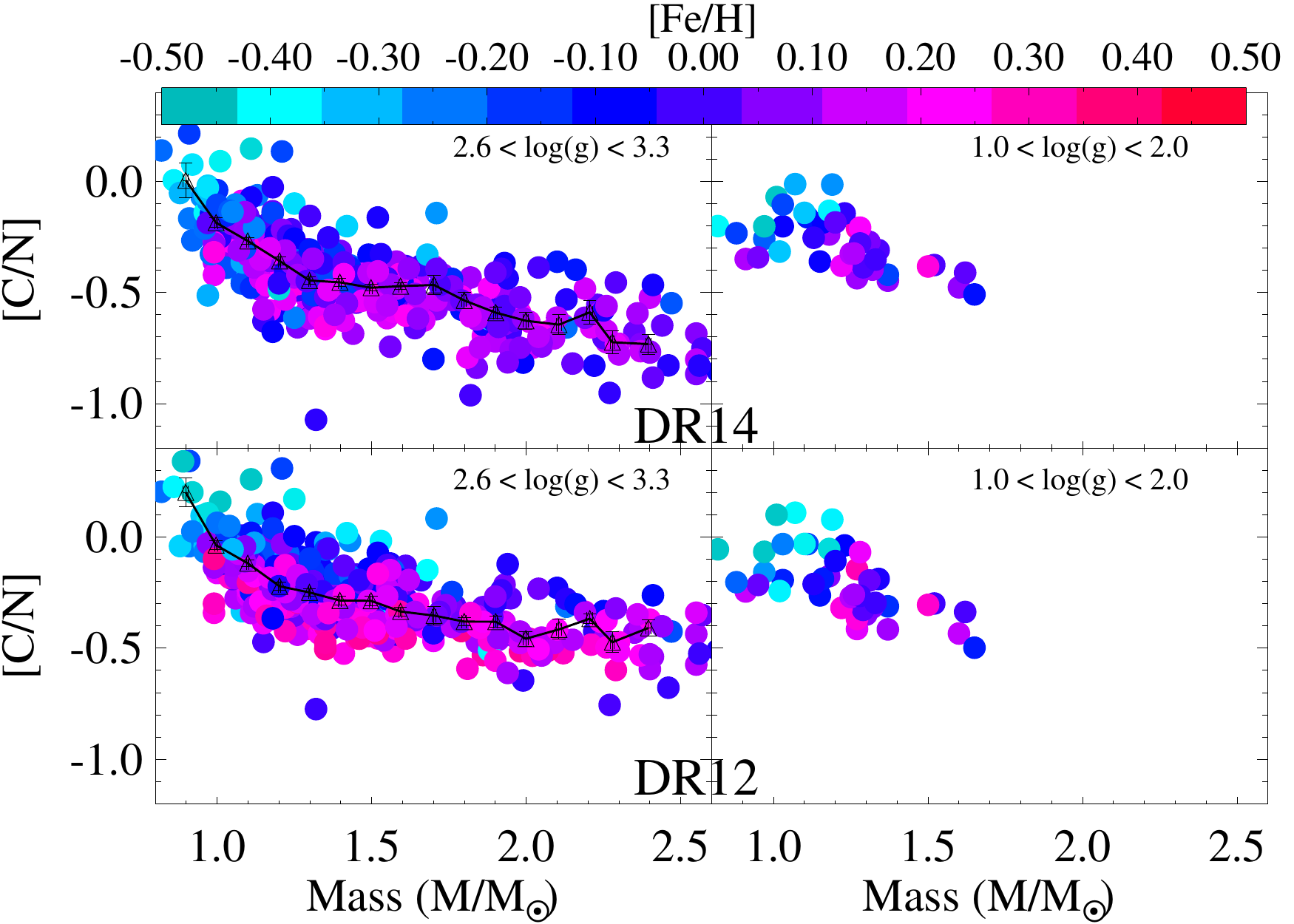} 
	\caption{[C/N] plotted as a function of seismic mass for both the LGB sample (left) and UGB sample (right). DR14 data are shown in the top row and DR12 data are shown in the bottom row. Points are colored by [Fe/H]. Median [C/N] values binned in mass are plotted as black open triangles.}
	\label{fig:martig_mass_plot_switch}
\end{figure*}

The APOGEE DR14 [C/N]-mass relation is more cleanly seen in Figure \ref{fig:martig_mass_plot_switch}, where we plot [C/N] vs. APOKASC seismic mass for the LGB (left) and UGB (right) samples described above. The points are colored by [Fe/H].  The LGB stars exhibit a clear correlation between [C/N] and mass for both DR14 and DR12 from 0.8 $<$ $M/M_{\odot}$ $<$ 2.0, indicating that the new DR14 [C/N] abundances can be used as mass indicators for these stars. However, the DR14 [C/N] and [Fe/H] values are slightly lower than the DR12 values (by $\sim$ 0.2 and $\sim$0.05 dex, respectively). The relation is relatively weak from 1.5 $<$ $M/M_{\odot}$ $< 2.5$, suggesting that [C/N] is not a very precise mass indicator for stars of these masses (see \citealt{Martig2016a} for an in-depth discussion of the [C/N]-mass precision). However, in \S \ref{sec:res}, where we present the results, we are most interested in comparing stars that are either ``young'', ``intermediate'', or ``old'' in age, which we can obtain from [C/N].
 
For the UGB sample, the parameter space is less-sampled, with only two stars with log(g) $<$ 1.5 and no stars with M $>$ 1.7 $M_{\odot}$. Therefore, there is insufficient data to determine whether DR14 [C/N] can be used as a mass indicator for the more luminous red giant stars with log(g) $<$ 1.5. However, because the stars that are found in this parameter space do show a relation between [C/N] and mass, we present the [C/N] results for these stars and accompanying age interpretations, as they are valuable tracers of the Galaxy at distances greater than 3 kpc from the Sun. Future studies of potential mixing mechanisms, as well as additional APOKASC observations, will result in a more concrete understanding of whether the [C/N] abundance of stars on the upper giant branch are indicative of their masses. 

\subsection{Mass to Age}
\label{sec:masstoage}

\begin{figure*}
\epsscale{1.0}
\plotone{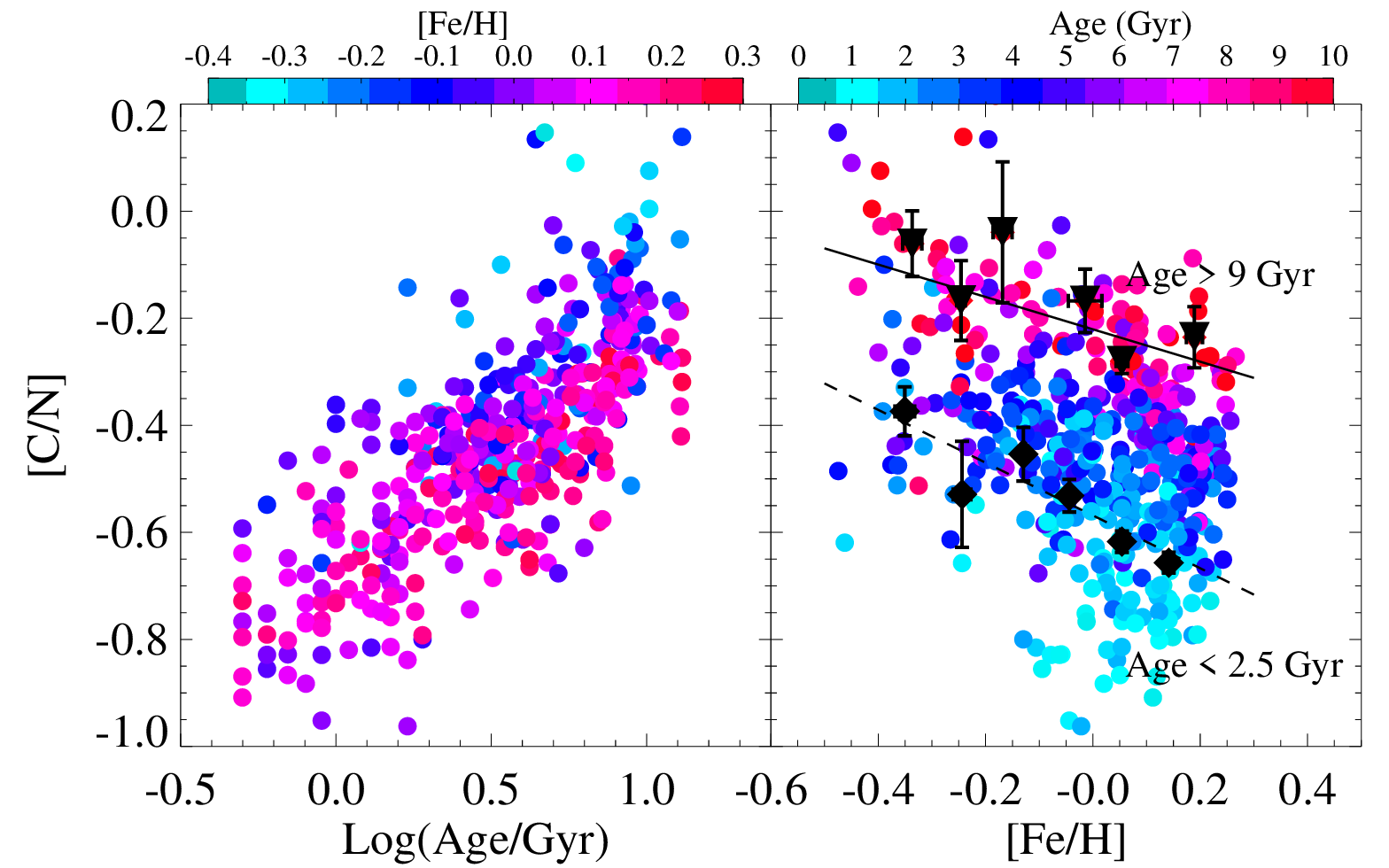} 
\caption{ Left: [C/N] vs. Log(Age/Gyr) from \citet{Martig2016a} for the APOKASC LGB sample. Points are colored by [Fe/H]. Right: [C/N] vs. [Fe/H] abundance tracks for the same stars in the left panel, colored by the input ages derived by \citet{Martig2016a}. The two lines indicate fits to median [C/N] abundances binned by [Fe/H] for stars with asteroseismic ages younger than 2.5 Gyr (black diamonds and dashed line) and older than 9 Gyr (black triangles and solid line).}
\label{fig:age_relate}
\end{figure*}

In this work we interpret [C/N]-[Fe/H] abundance trends not as mass-[Fe/H] trends, but as age-[Fe/H] trends. The age of a red giant star can be derived from its mass by invoking models of stellar evolution, as was done in \citet{Martig2016a}. In the left panel of Figure \ref{fig:age_relate}, we plot [C/N] vs. Log(Age/Gyr) derived from the seismic masses by \citet{Martig2016a} for the APOKASC LGB sample. This plot verifies the sensitivity of [C/N] abundance to stellar age. However, for Log(Age/Gyr) $>$ 0.3, there appears to be a slight color gradient, such that, at fixed age, the metal-poor stars tend to have higher [C/N] than the metal-rich stars. This is because the metallicity of a red giant star affects its age determination, as a more metal-poor star evolves faster through the main-sequence phase than a more metal-rich star of the same mass. This must be taken into account when interpreting the [C/N]-[Fe/H] abundance space as an age-abundance space.

In the right panel of Figure \ref{fig:age_relate}, we plot the [C/N]-[Fe/H] abundance space for the APOKASC LGB sample, colored by the age derived by \citet{Martig2016a} from the seismic masses. We divide the sample into a ``Young'' population, with age $<$ 2.5 Gyr, and an ``Old'' population with age $>$ 9 Gyr. We bin each sample into 0.1 dex bins of [Fe/H] and fit lines to the median [C/N] values of each bin. These lines show that stars of the same age follow sloped lines in the [C/N]-[Fe/H] abundance plane, which must be taken into consideration when comparing the relative ages of stars at different metallicities using [C/N] as an age indicator. In the following presentation of results, we over-plot these age tracks on the [C/N]-[Fe/H] abundance plots presented in \S \ref{sec:res}, which we use to guide our interpretation.

\section{Results}
\label{sec:res}

We present the  [C/N]-[Fe/H] abundance patterns for the MW Galaxy separately for the LGB and UGB sample. We divide the MW into 6 radial bins of 2 kpc each from 3-15 kpc. We also divide the samples according to height above the plane: 0 kpc $<$ $|$Z$|$ $<$ 0.5 kpc,  0.5 kpc $<$ $|$Z$|$ $<$ 1.0 kpc, 1.0 kpc $<$ $|$Z$|$ $<$ 2.0 kpc. In Figures \ref{fig:LGB} and \ref{fig:UGB}, the points are color-coded by $\alpha$-element abundance ([$\alpha$/Fe]) to explore potential [C/N] differences between the $\alpha$-element enhanced and solar-$\alpha$ MW populations, the former of which is generally thought to be older. Medians of [C/N] binned in 0.1 dex bins of [Fe/H] are plotted as black squares.

\subsection{LGB Sample}
\label{sec:LGB}

\begin{figure*}
\epsscale{1.0}
\plotone{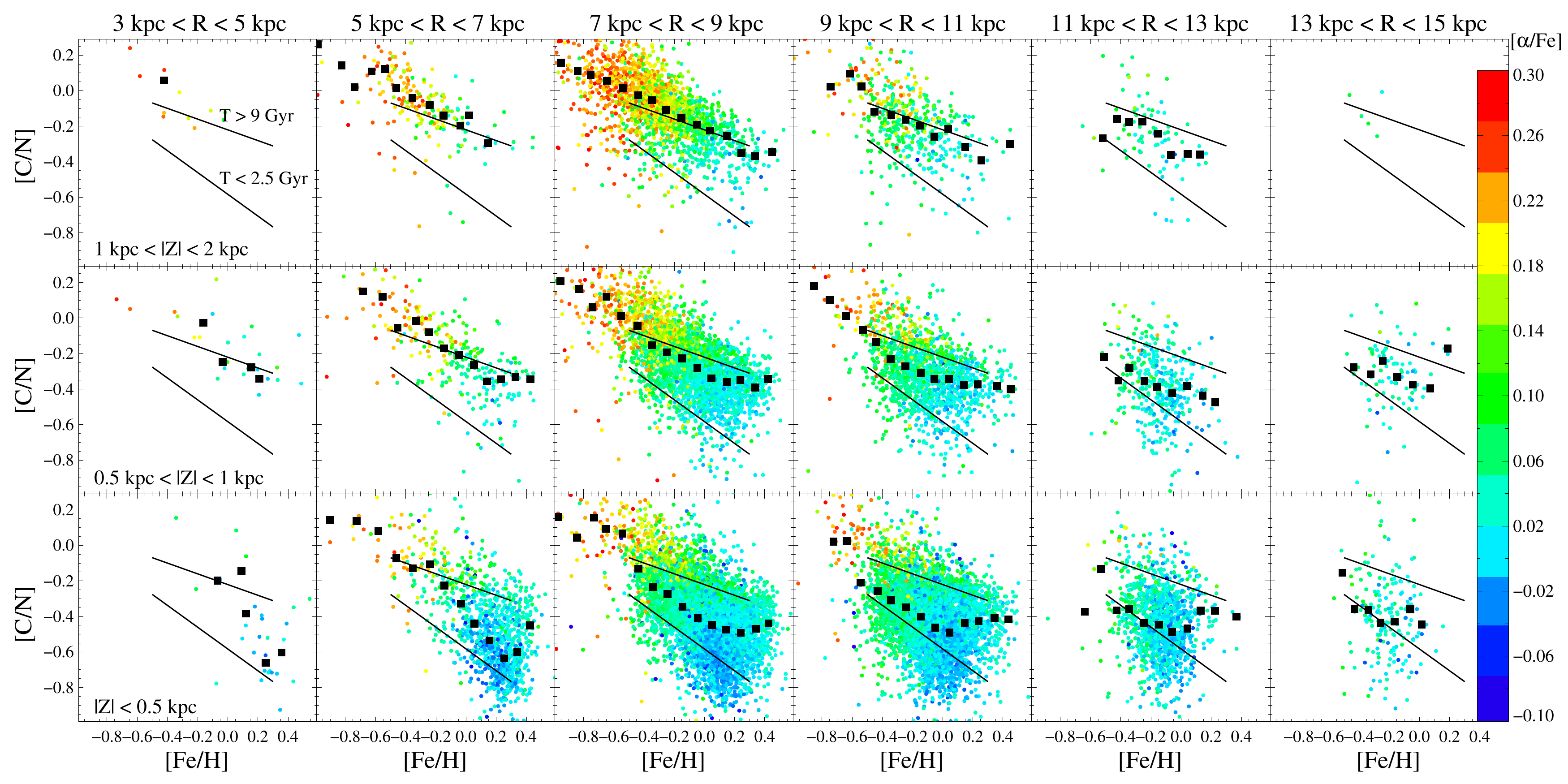} 
\caption{ [C/N] vs. [Fe/H] abundance patterns for the lower giant branch sample (LGB, 2.6 $<$ log(g) $<$ 3.3) across the Galaxy. Points are colored by [$\alpha$/Fe] and median [C/N] values in bins of [Fe/H] are plotted as black squares. Lines show the tracks for old (age $>$ 9 Gyr) and young (age $<$ 2.5 Gyr) stars from Figure \ref{fig:age_relate}.}
\label{fig:LGB}
\end{figure*}

The [C/N]-[Fe/H] abundance patterns for the LGB sample are plotted in Figure \ref{fig:LGB}. In each panel, the median [C/N]-[Fe/H] lines corresponding to ``Young'' and ``Old'' stars computed in \S \ref{sec:masstoage} are over-plotted as black lines. Because the LGB sample comprises the less luminous giant stars, Galactic coverage is largely limited to 5-13 kpc. 

%\begin{sidewaysfigure*}[ht]
         %\centering
	%\epsscale{1.0}
%	\includegraphics[width=\textwidth,height=\textheight,keepaspectratio]{LGB.pdf} 
	%\rule{0.75\textheight}{0.5\textheight}
%	\caption{[C/N] vs. [Fe/H] abundance patterns for the lower giant branch sample (LGB, 2.6 $<$ log(g) $<$ 3.3) across the Galaxy. Points are colored by [$\alpha$/Fe] and median [C/N] values in bins of [Fe/H] are plotted as black squares. Lines show the tracks for old (age $>$ 9 Gyr) and young (age $<$ 2.5 Gyr) stars from Figure \ref{fig:age_relate}.}
%	\label{fig:LGB}
%\end{sidewaysfigure*}

\subsubsection{Vertical Trends}

The stars with [Fe/H] $>$ $-$0.4 and 1 kpc $<$ $|$Z$|$ $<$ 2 kpc have higher [C/N] values than stars in the plane, indicating that these red giant stars are less massive and thus older, on average, than stars in the plane. There also appears to be a slight gradient in mean age from the inner Galaxy to the outer Galaxy for these stars, as the median [C/N] for stars with [Fe/H] $\sim$ $-$0.2 decreases from $\sim$ $-$0.05 dex in the 5 kpc $<$ R $<$ 7 kpc bin to $\sim$ $-$0.30 dex in the 11 kpc $<$ R $<$ 13 kpc bin. This is qualitatively consistent with the age gradient of $\sim$ $-$0.6 Gyr/kpc found for the thick disk by \citet{Martig2016b} using DR12 data, and in line with predictions by \citet{Minchev2015}, who suggest that the thick disk is a result of an inner disk born hot in a turbulent gas-rich phase, along with the flaring of mono-age stellar populations (see also \citealt{Mackereth2017}).

The stars that fall in the 1 kpc $<$ $|$Z$|$ $<$ 2 kpc and 7 kpc $<$ R $<$ 9 kpc bin appear to track the ``Old'' age line which corresponds to a median age of $\sim$ 10 Gyr. As has been shown in the APOGEE data by \citet{Hayden2015}, and observed in these data through the coloring of the points, this region of the Galaxy primarily consists of stars enhanced in the $\alpha$-elements. The [C/N] abundance trend seems to suggest that the ages of the stars at [Fe/H] = $-$0.7 do not differ significantly from the stars with [Fe/H] = +0.2. This indicates that the evolution of the gas from [Fe/H] = $-$0.7 to [Fe/H] = +0.2 happened over a relatively rapid timescale. 

\begin{figure*}
\epsscale{1.0}
\plotone{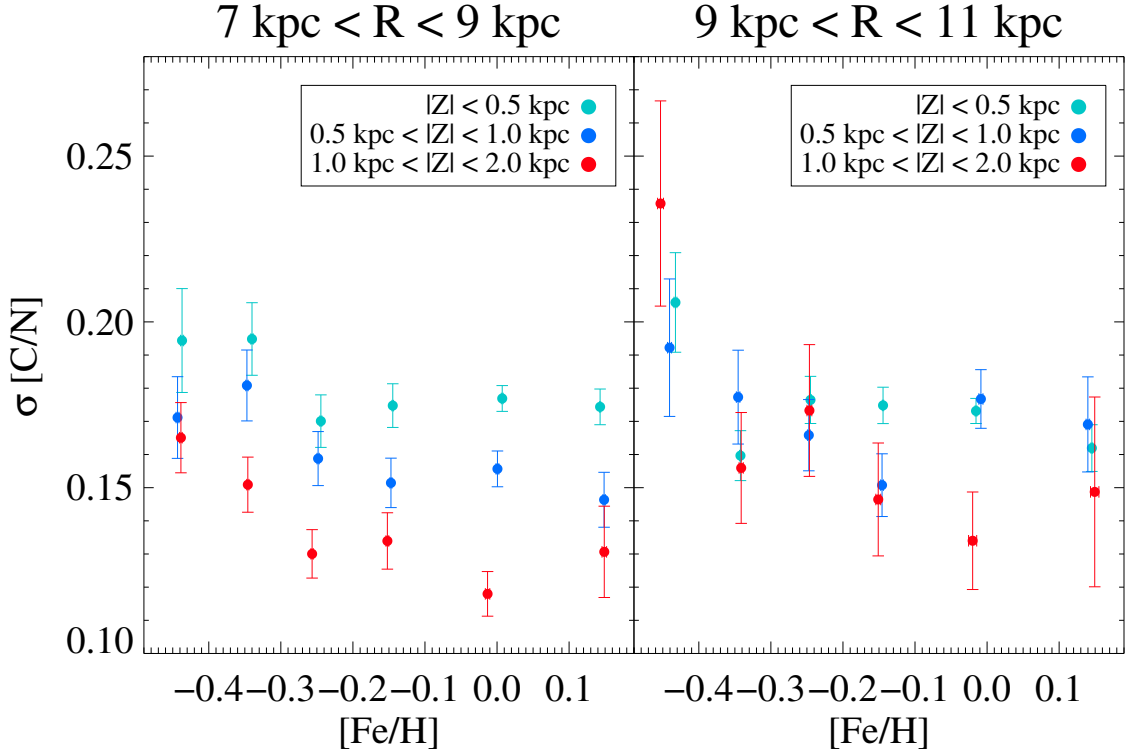} 
\caption{The dispersion of [C/N] binned in [Fe/H] for three different heights above the plane for stars with 7 kpc $<$ R $<$ 9 kpc (left) and 9 kpc $<$ R $<$ 11 kpc (right). Points are colored by $|$Z$|$, as indicated in the legend. }
\label{fig:sigma_cn}
\end{figure*}

While the median [C/N] indicates a relatively constant age with [Fe/H], there is significant dispersion (greater than the 0.07 dex measurement uncertainty) in [C/N] at each metallicity. This is quantified in Figure \ref{fig:sigma_cn}, where we plot the standard deviation of [C/N] in 0.1 dex bins of [Fe/H] for stars around the solar radius (left panel) and just outside the solar radius (right panel) for each $|$Z$|$ bin. At the solar radius, the [C/N] dispersion for stars with [Fe/H] $>$ $-$0.3 increases from 0.13 dex to 0.17 dex from out of the plane to in the plane. For stars with [Fe/H] $<$ $-$0.3, the dispersion is larger overall, but still increases from 0.16 to 0.19 dex from out of the plane to in the plane. The larger scatter in the plane is a consequence of the presence of young, low-[C/N] stars at $-$0.3 $<$ [Fe/H] $<$ 0.0, which are not found at $|$Z$|$ $>$ 1 kpc.

In the outer disk, the trend of increasing [C/N] dispersion with decreasing $|$Z$|$ is less-pronounced, as shown in the right panel of Figure \ref{fig:sigma_cn}. In fact, only stars in the [Fe/H] = 0.0 bin show a smaller dispersion of [C/N] out of the plane than in the plane, with all other bins exhibiting similar [C/N] dispersion across all $|$Z|$|$. The lack of variation is likely due, in part, to the fact that there are relatively few old, high-$\alpha$ stars populating the disk at R $>$ 9 kpc (see \citealt{Nidever2014,Hayden2015}), which exhibited tighter [C/N] dispersion above the plane at the Solar Neighborhood.

\subsubsection{Midplane Radial Trends}
\label{sec:LGB_plane}

The median [C/N] in each radial bin is lower for stars in the plane than stars out of the plane by $\sim$ 0.2 dex, indicating that stars in the plane are generally younger. This is seen in the data through a comparison of the top row of Figure \ref{fig:LGB} to the bottom row. The $\alpha$-element-enhanced stars in the plane generally have higher [C/N] than the solar-$\alpha$ abundance stars in the plane by $\sim$ 0.3 dex at a given [Fe/H], consistent with the $\alpha$-element-enhanced stars with $|$Z$|$ $>$ 1 kpc. Each radial bin in the plane (bottom row of Figure \ref{fig:LGB}) contains stars with [C/N] abundance values that fall $>$ 0.2 dex below the ``Young'' age line (age $<$ 2 Gyr). These stars tend to be deficient in [$\alpha$/Fe] by $\sim$ 0.1 dex relative to stars with larger [C/N] abundances at the same [Fe/H]. This correlation between $\alpha$-element abundance and [C/N] abundance has been observed in the APOGEE sample before (see, e.g., \citealt{Masseron&Gilmore2015,Ness2016,Martig2016a}). However, stars with $-$0.4 $<$ [Fe/H] $<$ $-$0.2 in the Solar Neighborhood exhibit a large spread in [C/N] such that some low-$\alpha$ stars have the same [C/N] abundance as the high-$\alpha$ stars, as can be seen by the coloring of points in Figure \ref{fig:LGB} in the Solar Neighborhood bin. Therefore, not only is [Fe/H] a poor tracer of age, as seen by the large spread in [C/N] abundance at a given [Fe/H] abundance, but stars with different $\alpha$-element abundances at a given [Fe/H] can be the same age, as suggested by \citet{Mackereth2018}. 

\begin{figure*}
\epsscale{1.0}
\plotone{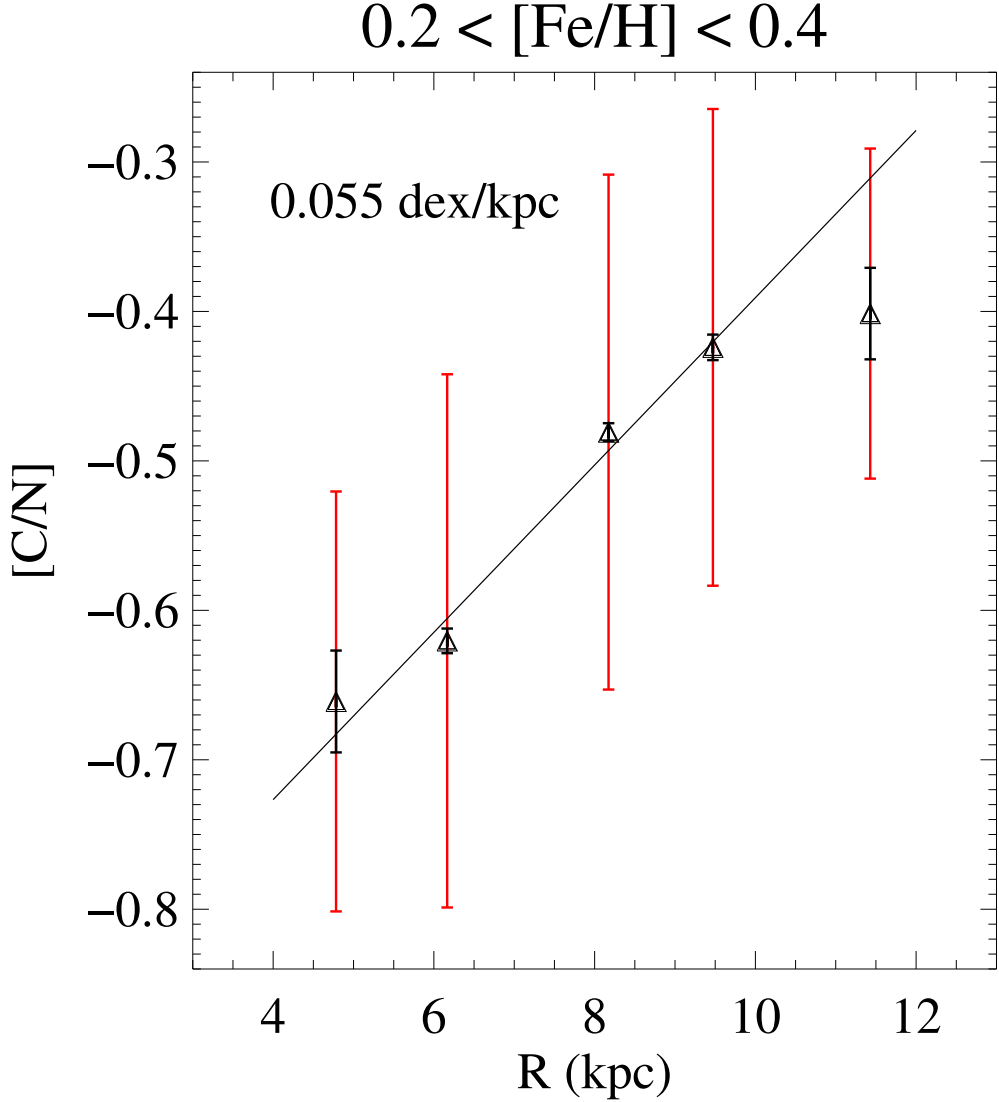} 
\caption{Median [C/N] values binned by Galactic radius for stars with +0.2 $<$ [Fe/H] $<$ +0.4 for the LGB sample. Black error bars indicate errors on the mean, and red error bars indicate the standard deviation. The errors on the mean are generally quite small ($<$ 0.02 dex), and therefore are not always visible in the plot.}
\label{fig:cn_grad}
\end{figure*}

In the plane, we observe [C/N]-[Fe/H] abundance trends with radius that are not strictly a result of the changing relative amounts of high-$\alpha$ and low-$\alpha$ populations. In the 5 kpc $<$ R $<$ 7 kpc, $|$Z$|$ < 0.5 kpc bin of Figure \ref{fig:LGB}, we observe a smooth decrease of [C/N] with increasing [Fe/H] from [Fe/H] = $-$0.6 to [Fe/H] = +0.2. At +0.3 $<$ [Fe/H] $<$ +0.5, there is an apparent \emph{increase} in median [C/N] from [C/N] = $-$0.6 to $-$0.4. We observe such an inflection point in each radial bin (bottom row of Figure \ref{fig:LGB}), but the inflection point (where [C/N] begins to increase rather than decrease) shifts to lower [Fe/H] with increasing radius.  In addition to the inflection point becoming more metal-poor, the median [C/N] abundance of the stars with +0.2 $<$ [Fe/H] $<$ +0.4 increases with increasing Galactic radius. This is shown in Figure \ref{fig:cn_grad} where the [C/N] increases from $-$0.6 dex to $-$0.4 dex from 5 kpc to 11 kpc. These trends suggest that at larger Galactic radii (R $>$ 9kpc), the metal-rich stars become older than at smaller radii, and can potentially be older than some of the more metal-poor stars in the same radial bin.

\begin{figure*}
\epsscale{1.0}
\plotone{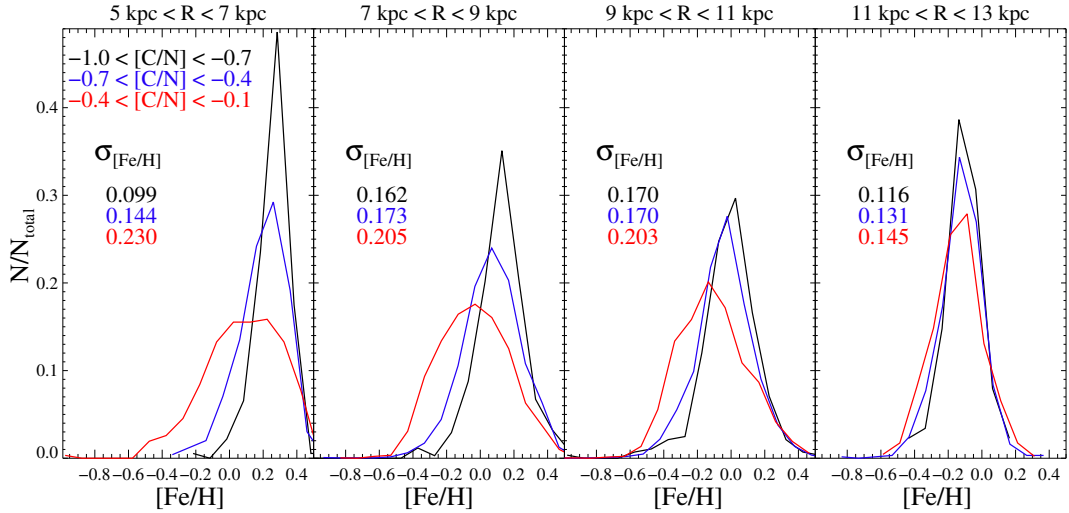} 
\caption{MDFs binned by [C/N] for four radial bins for stars with $|$Z$|$ $<$ 0.5 kpc. The MDFs are colored according to the [C/N] bin, as indicated in the top-left of the left panel, and are normalized to the total number of stars in each bin. The standard deviation of [Fe/H] in each bin is printed in each panel and colored according to [C/N] bin.}
\label{fig:mdf_bin}
\end{figure*}

The stars with the lowest [C/N] abundance in each radial bin in the plane (bottom row of Figure \ref{fig:LGB}), which likely represent the stars formed in the most recent epoch of star formation in that radial bin, exhibit an apparent anti-correlation between mean [Fe/H] abundance and Galactic radius, indicating there is a negative radial metallicity gradient in the gas participating in star formation in the Galaxy. Additionally, with the exception of the 11 kpc $<$ R $<$ 13 kpc bin, these lowest [C/N], young stars generally exhibit a narrower MDF than stars of larger [C/N] in the same radial bin, as shown in Figure \ref{fig:mdf_bin}, where the MDFs in bins of [C/N] are plotted for the $|$Z$|$ $<$ 0.5 kpc radial bins. The widths of the MDFs are estimated as the standard deviation of [Fe/H] in each bin, and are listed in each panel of Fiugre \ref{fig:mdf_bin}. These observations are consistent with MW formation scenarios in which there was a metallicity gradient in the gas participating in star formation across the Galaxy, and this star formation occurs today with gas at one metallicity rather than spanning a range of metallicities. The lack of MDF evolution with [C/N] in the outer Galaxy is likely in part due to the lack of more metal-poor (right-most panel of Figure \ref{fig:mdf_bin}), $\alpha$-enhanced stars that are present at R $<$ 9 kpc (see, e.g., \citealt{Hayden2015}). We explore further interpretations in \S \ref{sec:mig}.

\begin{figure*}
\epsscale{1.0}
\plotone{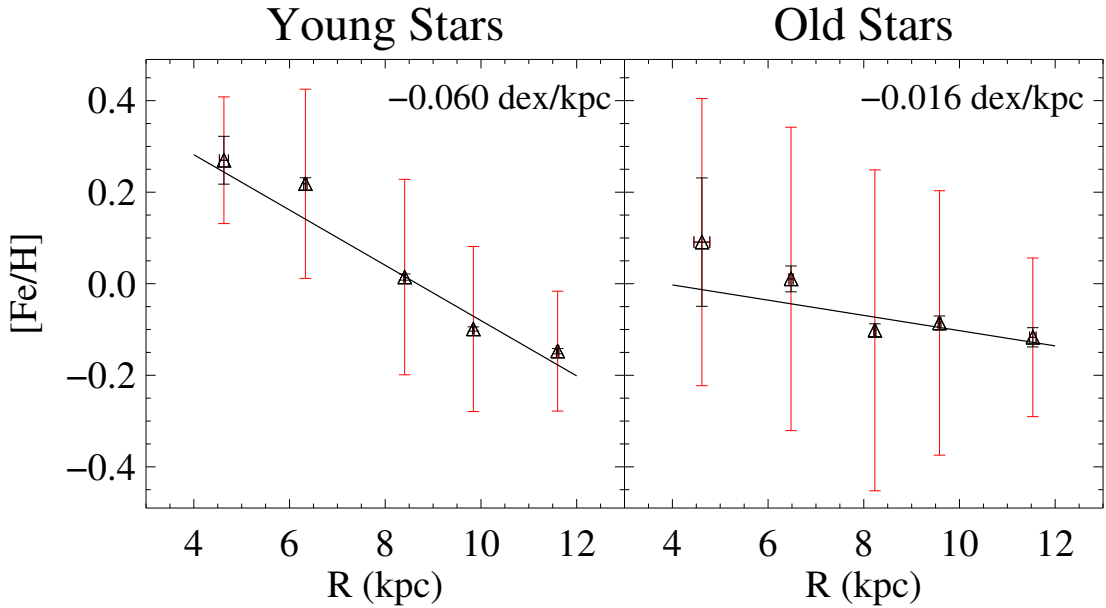} 
\caption{Left: Median [Fe/H] values binned by Galactic radius for Young Stars as described in the text. Right: Median [Fe/H] values binned by Galactic radius for Old Stars as described in the text. Black error bars indicate errors on the mean and red error bars indicate standard deviation. The errors on the mean are generally quite small ($<$ 0.02 dex), and therefore are not always visible in the plot.}
\label{fig:grad}
\end{figure*}

To quantify the metallicity gradient of the low-[C/N] stars, we plot the median [Fe/H] for the stars with lower [C/N] abundances than the ``Young'' age track as a function of Galactic radius, shown in the left panel of Figure \ref{fig:grad}.  We fit a line to the medians, weighted by error on the mean, and find that there is a gradient in [Fe/H] of $-$0.060 $\pm$ 0.002 dex/kpc for these young stars from 5-12 kpc. \citet{Anders2017b} found a gradient of $-$0.057 dex/kpc for the CoRoGEE stars with ages less than 1 Gyr and $-$0.066 dex/kpc for the CoRoGEE stars with ages between 1 and 2 Gyr. The gradient found here is also in good agreement with the \citet{Genovali2014} gradient of  $-$0.06 dex/kpc derived from Cepheid stars. It is considerably steeper than work by \citet{Balser2011}, who find gradients $\sim$ $-$0.03-0.04 dex/kpc using metallicities measured from H II regions, but some as high as $-$0.07 dex/kpc, depending on azimuthal angle. It is also steeper than the gradient found for young planetary nebula by \citet{Stanghellini2018} of $-$0.027 dex/kpc. 

Similar gradient analysis with old stars is shown in the right panel of Figure \ref{fig:grad}. For the older stars we find a flatter gradient of $-$0.016 $\pm$ 0.005 dex/kpc. This is consistent with what \citet{Anders2017b} finds for CoRoGEE stars older than 10 Gyr ($-$0.021 dex/kpc). This is flatter than the gradients found for old open clusters in the APOGEE data (see, e.g., \citealt{Cunha2016}), but is consistent with old planetary nebulae measurements from \citet{Stanghellini2018} of $-$0.015 dex/kpc. A flattening of the gradient for older stellar populations is predicted by models of Galaxy evolution in which stars migrate over time (see, e.g., \citealt{Minchev2014}).

\subsection{UGB Sample}

\begin{figure*}
\epsscale{1.0}
\plotone{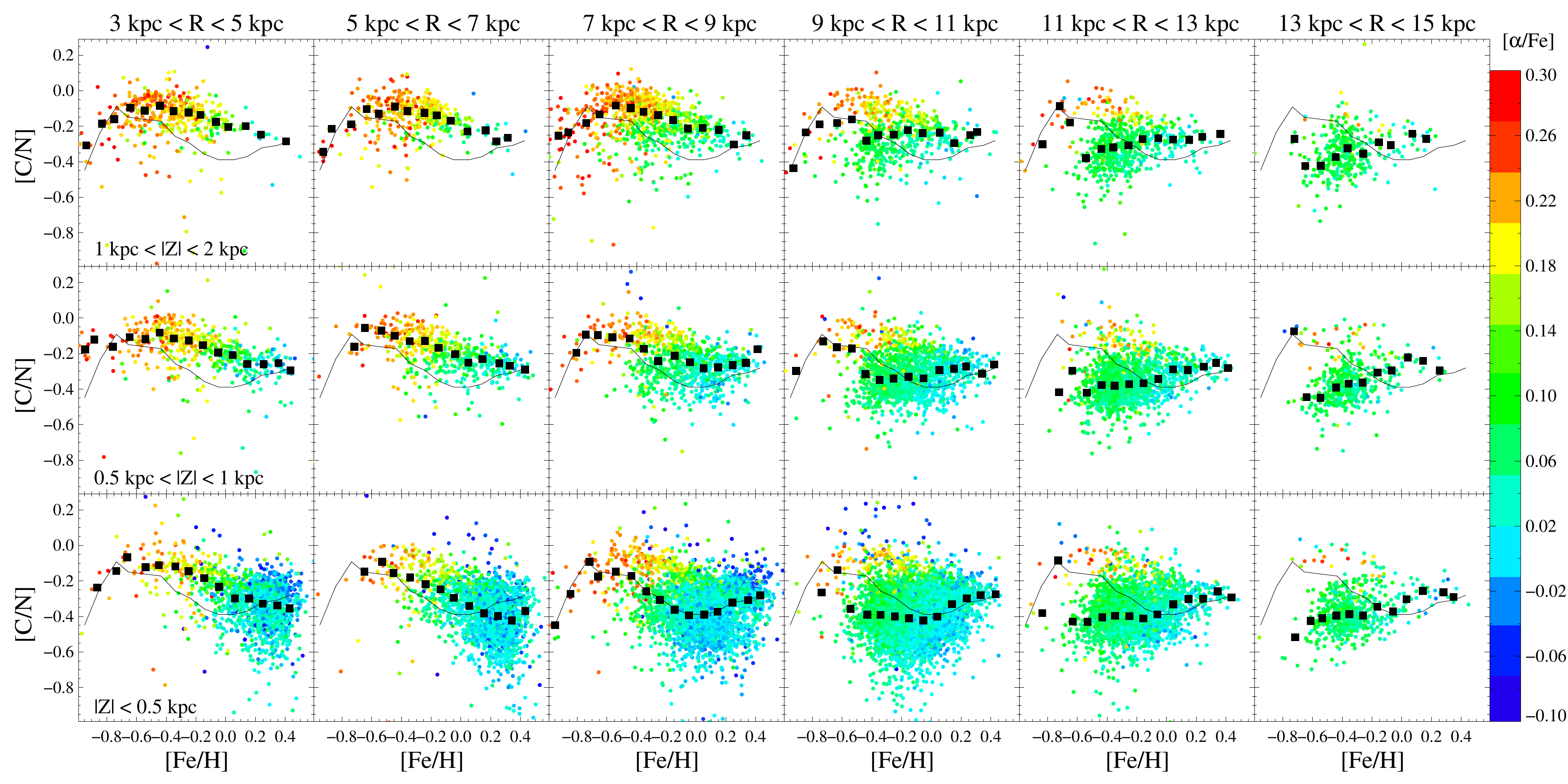} 
\caption{Similar to Figure \ref{fig:LGB} but for the upper giant branch (UGB) stars (1.0 $<$ log(g) $<$ 2.1). Median [C/N] values in bins of [Fe/H] are plotted as black squares. The black line in all panels is a reproduction of the median [C/N] values in bins of [Fe/H] from the 7 kpc $<$ R $<$ 9 kpc, $|$Z$|$ $<$ 0.5 kpc bin.}
\label{fig:UGB}
\end{figure*}

The UGB sample traces larger Galactic distances than the LGB sample. However, as explained in \S \ref{sec:sample}, there is a lack of APOKASC stars with log(g) $<$ 2.0 that have seismic masses. Additionally, there are extra-mixing effects that are a function of metallicity and log(g), which can alter the surface [C/N] abundances after the initial, mass-dependent dredge up (see, e.g., \citealt{Gilroy1989,Gratton2000,Lagarde2017,Lagarde2018}). Therefore, it is not clear whether or not the [C/N] abundances for these stars can indicate their age. However, despite these added complications in using [C/N] as a mass indicator, Figure \ref{fig:UGB} shows that the UGB sample exhibits similar [C/N]-[Fe/H] abundance patterns to the LGB sample where they overlap spatially. In these plots, rather than showing age tracks, we over-plot the median [C/N]-[Fe/H] abundance pattern from the 7 kpc $<$ R $<$ 9 kpc, $|$Z$|$ $<$ 0.5 kpc bin in all other bins so that we can compare other regions of the Galaxy to the Solar Neighborhood. In the following sections, we analyze regions of the Galaxy not probed by the LGB sample, as well as highlight instances where the LGB and UGB [C/N]-[Fe/H] abundances diverge.

%\begin{sidewaysfigure*}
%\epsscale{1.0}
%\includegraphics[width=\textwidth]{UGBE.pdf} 
%\caption{Similar to Figure \ref{fig:LGB} but for the upper giant branch (UGB) stars (1.0 $<$ log(g) $<$ 2.1). Median [C/N] values in bins of [Fe/H] are plotted as black squares. The black line in all panels is a reproduction of the median [C/N] values in bins of [Fe/H] from the 7 kpc $<$ R $<$ 9 kpc, $|$Z$|$ $<$ 0.5 kpc bin.}
%\label{fig:UGB}
%\end{sidewaysfigure*}

\subsubsection{Vertical Trends}
The UGB sample extends to the 3 kpc $<$ R $<$ 5 kpc and 13 kpc $<$ R $<$ 15 kpc bins, which were hardly probed by the LGB sample. Far from the plane (1 kpc $<$ $|$Z$|$ $<$ 2 kpc), the innermost bin shows a similar [C/N]-[Fe/H] abundance trend as the 5 kpc $<$ R$<$ 7 kpc bin. The outermost bin follows a continuation of the [C/N] gradient where the median [C/N] abundance falls from $\sim$ $-$0.15 dex to $\sim$ $-$0.35 dex from 5 kpc to 15 kpc.

The qualitatively similar behavior of the [C/N]-[Fe/H] abundance tracks for the UGB sample at [Fe/H] $>$ $-$0.5 indicates that there is likely still mass information in the [C/N] abundances for these stars. However, at [Fe/H] $<$ $-$0.5, the [C/N]-[Fe/H] abundance tracks diverge from the LGB sample. The LGB sample showed an anti-correlation between [C/N] and [Fe/H] across the full metallicity range whereas the UGB sample begins to show a \emph{correlation} between [C/N] and [Fe/H] at [Fe/H] $<$ $-$0.5. This is likely a sign of extra mixing along the giant branch that begins at [Fe/H] $<$ $-$0.5 (see, e.g., \citealt{Lagarde2017} and investigated in the APOGEE sample by \citealt{Shetrone2018:inprep}). Additionally, the range in [C/N] values for the UGB sample above the plane and in the 7 kpc $<$ R $<$ 9 kpc bin is $\sim$ 0.4 dex, whereas the range for the LGB sample at this same Galactic position is $\sim$ 0.7 dex. This suggests that extra mixing occurs along the upper giant branch in such a way that reduces the sensitivity of [C/N] to mass.

At R $>$ 11 kpc, which were regions of the Galaxy not sufficiently probed in the LGB sample, the stars above the plane no longer follow the [C/N]-[Fe/H] anti-correlation, and in fact, exhibit somewhat of a [C/N]-[Fe/H] correlation. As shown in \citet{Hayden2015}, and indicated by the coloring of the points in Figure \ref{fig:UGB}, the $\alpha$-enhanced stars at [Fe/H] $<$ $-$0.2 no longer occupy this region of the Galaxy. The stars that populate this region of the Galaxy are likely flared ``thin-disk'' stars. The solar-$\alpha$, [Fe/H] $<$ $-$0.2 stars in the outer Galaxy actually exhibit deficient [C/N] relative to the stars with [Fe/H] $>$ 0.0 in the same bin by $\sim$ 0.1 dex. This could be an indication that the metal-rich stars in the outer Galaxy are actually older than the more metal-poor stars, a feature that is more pronounced in stars that reside in the plane of the MW and seen in the LGB sample.

\subsubsection{Midplane Radial Trends}

\begin{figure*}
\epsscale{1.0}
\plotone{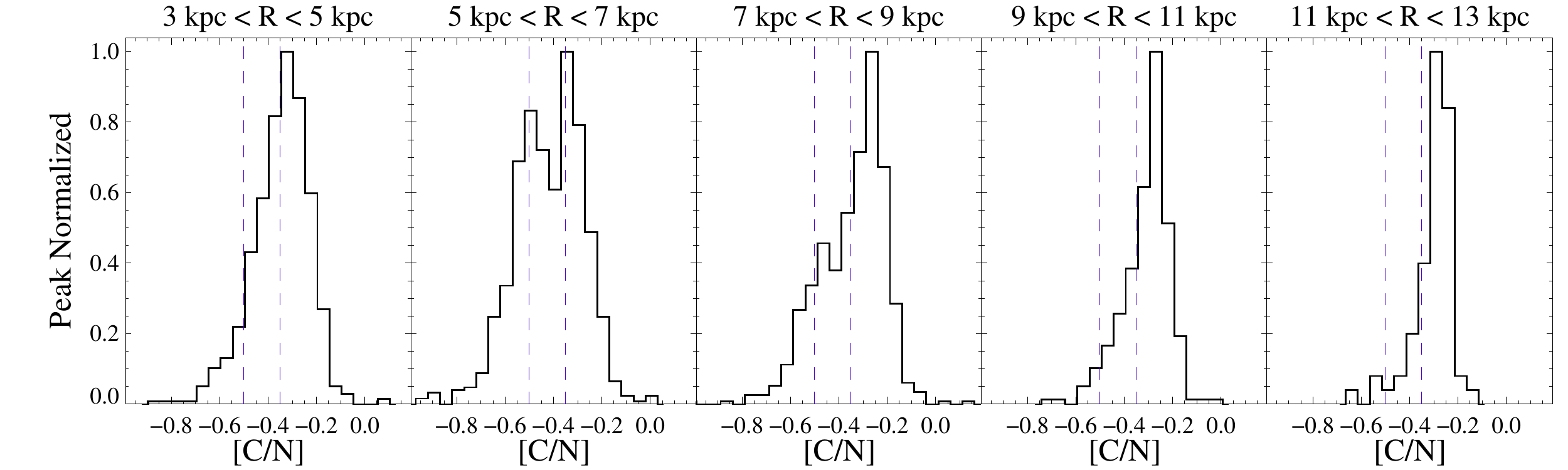} 
\caption{[C/N] histogram for the metal-rich UGB stars (1.0 $<$ log(g) $<$ 2.1 and 0.2 $<$ [Fe/H] $<$ 0.4). Only stars with $|$Z$|$ $<$ 0.5 kpc are plotted. Two fiducial lines meant to roughly track the peaks in the 5 kpc $<$ R $<$ 7 kpc bin are plotted in each panel.}
\label{fig:cn_hist}
\end{figure*}

In the plane, the 5 kpc $<$ R $<$ 7 kpc bin contains a greater fraction of metal-rich, low-[C/N] stars than the innermost 3 kpc $<$ R $<$ 5 kpc bin. This is better shown in Figure \ref{fig:cn_hist} where we plot [C/N] distribution functions for the stars with +0.2 $<$ [Fe/H] $<$ +0.4. The distribution in the 5 kpc $<$ R $<$ 7 kpc is double-peaked, with a peak centered around [C/N] = $-$0.5 that is hardly represented in the 3 kpc $<$ R $<$ 5 kpc bin. If we take [C/N] to indicate age for the UGB stars, this would suggest that the innermost regions of the Galaxy (R $<$ 5 kpc) do not have the youngest stars observed in other regions of the Galaxy (particularly the 5 kpc $<$ R $<$ 7 kpc bin), implying that there has been relatively low star formation in recent times as compared to stars with 5 kpc $<$ R $<$ 9 kpc.

%\begin{sidewaysfigure*}
%\epsscale{1.0}
%\includegraphics[width=\textwidth]{cn_rich_hist_UGB.eps} 
%\caption{[C/N] histogram for the metal-rich UGB stars (1.0 $<$ log(g) $<$ 2.1 and 0.2 $<$ [Fe/H] $<$ 0.4). Only stars with $|$Z$|$ $<$ 0.5 kpc are plotted. Two fiducial lines meant to roughly track the peaks in the 5 kpc $<$ R $<$ 7 kpc bin are plotted in each panel. }
%\label{fig:cn_hist}
%\end{sidewaysfigure*}

\section{Interpretations}
\label{sec:disc}

\subsection{Chemical Indication of Radial Migration}
\label{sec:mig}

Based on some Galaxy simulations (see, e.g., \citealt{Roskar2008,Bird2013,Minchev2013}), as well as observations of a change in skewness of the MDFs across the Galaxy presented in \citet{Hayden2015}, we expect to see some signature of radial migration in exploring the ages of stellar populations of the MW. \citet{Masseron&Gilmore2015} noticed that there appeared to be a group of super-solar metallicity stars in the APOGEE data with unexpectedly high [C/N]. One explanation for the existence of these metal-rich stars is that they actually formed in the inner Galaxy, where higher ISM metallicities were achieved at earlier times as compared to the outer Galaxy. Over time, via dynamical interactions with the bar and spiral arms, these stars migrated to their present location (see, e.g., \citealt{Minchev2013,Kordopatis2015}).  

There are two features of radial migration that might manifest themselves in the APOGEE [C/N]-[Fe/H] abundance patterns. First, radial migration scatters stars both ways. Because the outer disk is less-dense than the inner disk, the stars that migrate from in-to-out are expected to make a larger fractional contribution to the stellar populations of the outer disk than stars moving out-to-in (see, e.g., \citealt{Minchev2014}). Secondly, the older stars will migrate the farthest as they have had more time to migrate. The youngest stars should not have had time to migrate, implying that the metallicity of the youngest stars in a radial bin is a representation of the present-day ISM metallicity in that radial bin. If there are stars more metal-rich than this metallicity, they might be migrators. To search for signatures of radial migration in the APOGEE data, we look at [C/N] distributions in bins of [Fe/H] across the entire Galaxy. 

For each radial bin, we define a metallicity of most recent star formation, [Fe/H]$'$. [Fe/H]$'$ is defined as the median metallicity of the 20 stars with the lowest [C/N] in a given radial bin. For each radial bin, we select stars within 0.1 dex of [Fe/H]$'$ and analyze their [C/N] distributions. Without radial migration or the dilution of the ISM via accretion of a significant amount of pristine gas (explored more in \S \ref{sec:infall}), a simple prediction from chemical evolution would be that the [C/N] distribution of stars within 0.1 dex of [Fe/H]$'$ should be skewed heavily towards lower [C/N] values, or younger ages. In other words, the most-metal rich stars would be the youngest and there would be few old stars that have this same metallicity, as they would have had to form before the ISM was enriched to its current metallicity. 

\begin{figure*}
\epsscale{1.0}
\plotone{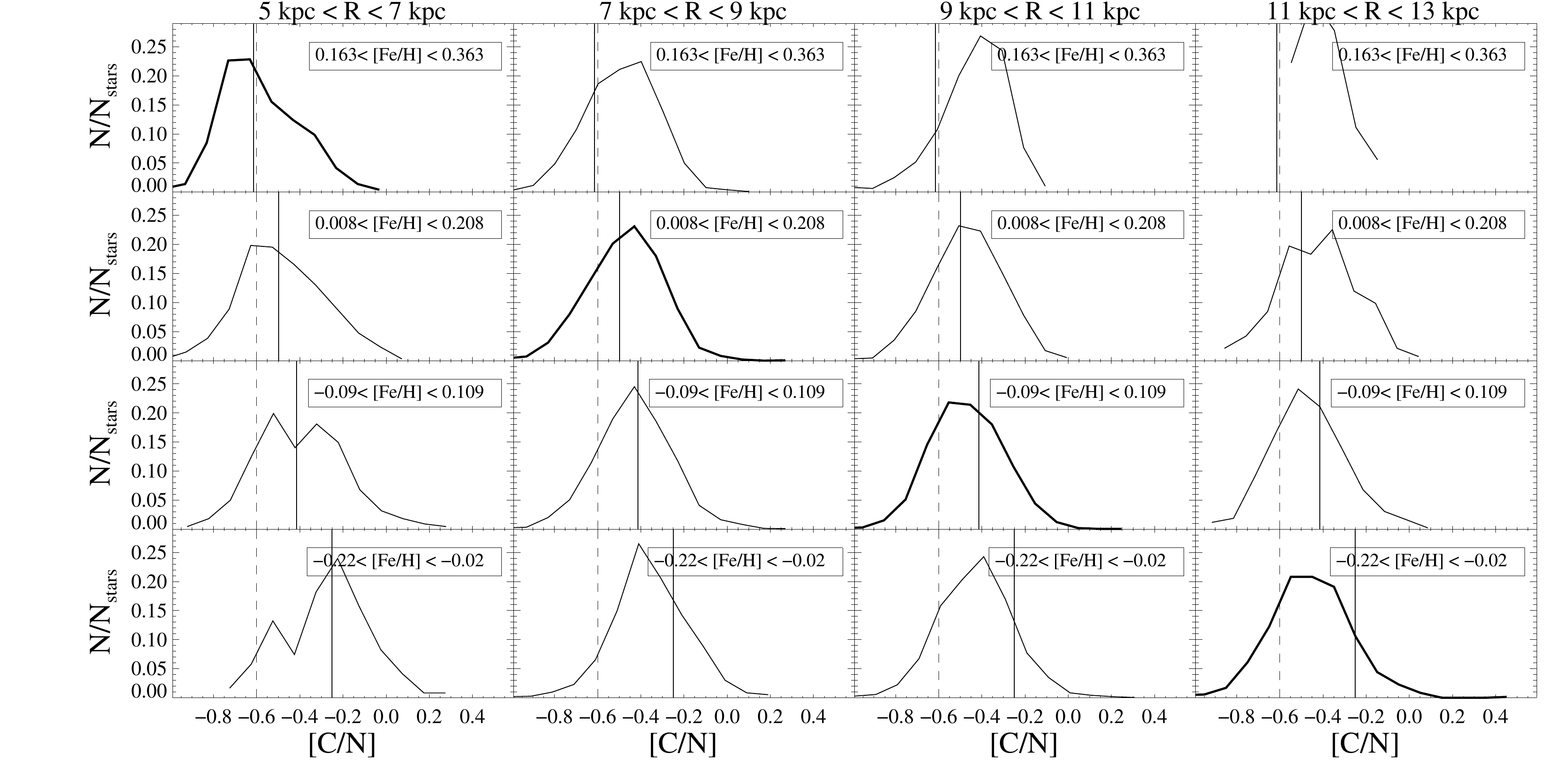} 
\caption{[C/N] distributions for stars $\pm 0.1$ dex from four different [Fe/H] values (indicated in the upper right of each panel) from the inner Galaxy to the outer Galaxy (columns, with radial range as marked).  The four [Fe/H] values correspond to the median metallicity of the most recently formed stars (specifically, of the 20 stars with the lowest [C/N]) in each radial bin. The solid line marks the median [C/N] of the inner Galaxy bin, and is the same across each row. The dashed line is a fiducial line at [C/N] = $-$0.6, which serves as an estimate for indicating young stars.}
\label{fig:LGB_spread}
\end{figure*}

The resulting [C/N] distributions for different values of [Fe/H]$'$ across \emph{all} radial bins are shown in Figure \ref{fig:LGB_spread}. In each panel, the [Fe/H]$'$ considered is shown in the upper right. The dashed line is a fiducial line at [C/N] = $-$0.6, which serves as an estimate for indicating young stars. The solid line is the median [C/N] value in the inner Galaxy, and is the same across each row. The top row shows the [C/N] distributions for the [Fe/H]$'$ of the inner Galaxy. At these metallicities, stars are skewed to younger ages in the inner Galaxy, but skewed to older ages in the outer Galaxy. This suggests that the metal-rich stars in the inner Galaxy that have just formed there dominate over older stars that potentially migrated from further in still, whereas the outer Galaxy only contains old metal-rich stars. These features are both consistent with radial migration. 

%\begin{sidewaysfigure*}
%	\epsscale{1.0}
%	\includegraphics[width=\textwidth]{LGB_age_spread.eps} 
%	\caption{[C/N] distributions for stars $\pm 0.1$ dex from four different [Fe/H] values (indicated in the upper right of each panel) from the inner Galaxy to the outer Galaxy (columns, with radial range as marked).  The four [Fe/H] values correspond to the median metallicity of the most recently formed stars (specifically, of the 20 stars with the lowest [C/N]) in each radial bin. The solid line marks the median [C/N] of the inner Galaxy bin, and is the same across each row. The dashed line is a fiducial line at [C/N] = $-$0.6, which serves as an estimate for indicating young stars. }
%	\label{fig:LGB_spread}
%\end{sidewaysfigure*}

The next row down in Figure \ref{fig:LGB_spread} shows the [C/N] distribution of stars with the [Fe/H]$'$ of the Solar Neighborhood. In the Solar Neighborhood bin, these stars are not skewed to younger ages. In fact, the distribution peaks at a larger value of [C/N], indicating that the majority of stars with the [Fe/H]$'$ of the Solar Neighborhood are actually more of an intermediate age, rather than skewed to younger ages as was the case for the inner Galaxy [Fe/H]$'$. This is qualitatively consistent with \citet{Minchev2014} who find that the majority of stars found in the Solar Neighborhood today were not formed there, with $\sim$ 60\% of the stars arriving from the inner disk. Assuming some of these stars arriving from the inner disk had metallicities similar to the Solar Neighborhood [Fe/H]$'$, this would explain the [C/N] distribution being peaked at larger values. We actually find that the median [C/N] of stars with Solar Neighborhood [Fe/H]$'$ peaks at roughly the same place ($\sim$ $-$0.45) across the entire Galaxy studied here. Similar patterns to the Solar Neighborhood [Fe/H]$'$ distributions are observed in the 9 kpc $<$ R $<$ 11 kpc bin (third row of Figure \ref{fig:LGB_spread}). 

The outermost radial bin shows that the metal-rich stars there are all old, but the stars with the metallicity of the outer Galaxy [Fe/H]$'$ (bottom row of Figure \ref{fig:LGB_spread}) exhibit a range of [C/N] values, suggesting a range of ages. This is consistent with radial migration in that the spread of this bin can be attributed to some intermediate age stars that formed in the next bin over (9 kpc $<$ R $<$ 11 kpc) migrating outwards over time. These stars appear to dominate over longer term migrators that may have come from the inner Galaxy at this lower [Fe/H]. 

This interpretation is all based on the assumption that the present day ISM at a given Galactic bin is as metal-rich as it has ever been, and that it was more metal-poor in the past. This may not be true for the Solar Neighborhood, where chemical evolution may have been slow in the past 4.5 Gyr (see e.g, \citealt{Chiappini2003}). Moreover, some chemical-evolution models show that the ISM can quickly evolve to an ``equilibrium abundance'', at which stars are formed over 6-8 Gyr (see, e.g., \citealt{Andrews2017,Weinberg2017}). This suggests that a spread in [C/N] at the [Fe/H]$'$ of a given radial bin is due to stars forming at this equilibrium abundance over several Gyr, rather than a sign of radial migration. Potential migrators would then be stars that are more metal-poor than the [Fe/H]$'$ of a given radial bin but younger (e.g., stars with low [C/N] in the [Fe/H] $<$ [Fe/H]$'$ panels of the inner Galaxy column in Figure \ref{fig:LGB_spread}). However, since these stars would not have had much time to migrate, they would not be able to stray too far from their birth location. Therefore, an equilibrium abundance scenario would have to explain the presence of metal-poor young stars in a radial bin that's been forming stars at a much higher metallicity for several Gyr, like we observe in Figure \ref{fig:LGB_spread}.  

\subsection{Primordial Variation}
\label{sec:AGB}

Interpretations presented in this paper are based on the assumption that the observed variation in [C/N] abundance is dominated by stellar evolution, and is not a reflection of primordial variations in the C and N abundances across the Galaxy. As mentioned earlier, analyses of the Solar Neighborhood have found some slight ($\sim$ 0.1 dex) variations in the [C/Fe] abundance of solar twins (see, e.g., \citealt{Nissen2015}), but it is unknown how much the primordial [C/N] abundance distribution varies across the Galaxy. However, there have been discoveries of chemically anomalous N-enhanced field stars in the APOGEE data (see, e.g., \citealt{Fernandez-Trincado2016,Fernandez-Trincado2017,Schiavon2017}), which might show up in our sample as low-[C/N] stars. These stars are few in number (11 found in the disk), and generally exhibit [Fe/H] $<$ $-$0.5, so their contribution to the gradient and migration analysis is likely insignificant.  

In addition to anomalous field stars being present, different star-formation histories of the inner and outer disk might result in variation of the primordial [C/N] abundance. For example, \citet{Bensby2005} found signs of $s$-process enrichment in the Galactic thin disk around the Solar Neighborhood, which suggests that AGB stars might be important contributors to the formation of these stars. In the inner Galaxy, where the chemical enrichment was more rapid, Type II SNe likely dominated the chemical enrichment without much time for AGB stars to contribute. If the AGB stars contribute additional C and/or N relative to Fe, then the metal-rich stars in the outer Galaxy may simply have been formed with a different primordial [C/N] abundance than the inner Galaxy. \citet{Masseron&Gilmore2015} found that the APOGEE data exhibited in increase in [(C+N)/Fe] with increasing [Fe/H] above solar metallicities, which they attributed to potential AGB enrichment. 

\begin{figure*}
\epsscale{1.0}
\plotone{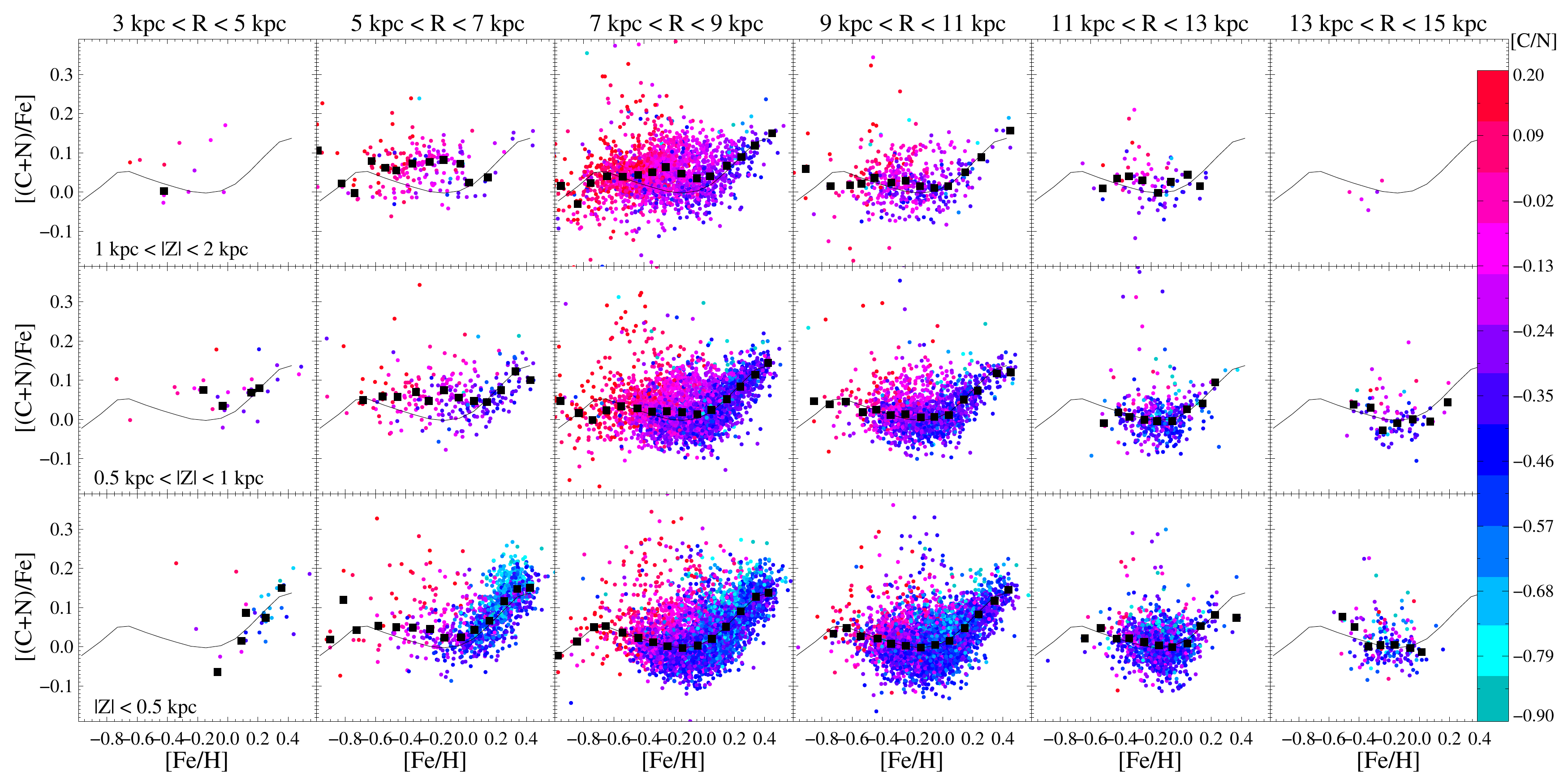} 
\caption{[(C+N)/Fe] for the lower giant branch stars (2.6 $<$ log(g) $<$ 3.3). Points are color-coded by [C/N] abundance. Medians binned in [Fe/H] are plotted as black squares. A fiducial line from the Solar Neighborhood, $|$Z$|$ $>$ 0.5 kpc bin is plotted in all panels.}
\label{fig:LGB_c+n}
\end{figure*}

In Figure \ref{fig:LGB_c+n} we plot the [(C+N)/Fe] vs. [Fe/H] abundance tracks for the LGB sample, this time colored by [C/N], to investigate potential differences in the [(C+N)/Fe] abundance patterns between the inner and outer Galaxy that may affect our [C/N]-age interpretations. We find that the stars with $|$Z$|$ $>$ 1 kpc have larger [(C+N)/Fe] abundances than the stars in the plane at [Fe/H] $\sim$ $-$0.2, but are mostly similar at other metallicities. This is akin to the $\alpha$-element abundance differences, and is consistent with the $\alpha$-element enhanced stars having formed with a larger relative amount of Type II SNe ejecta. At super-solar [Fe/H], we see the same increase of [(C+N)/Fe] with [Fe/H] that \citet{Masseron&Gilmore2015} observed, but we observe this trend in every radial/Z bin. The fiducial line from the Solar Neighborhood indicates that there is little variation in this trend between bins. This suggests that whatever process (or processes) causes an increase in [(C+N)/Fe] with increasing [Fe/H] at [Fe/H] $>$ 0.0 operates in a similar fashion between the inner Galaxy and outer Galaxy. We investigated the same tracks for the UGB sample and found similar results. All Galactic zones show an increase in [(C+N)/Fe] with increasing [Fe/H] at super-solar metallicities.

%\begin{sidewaysfigure*}[h]
%\epsscale{1.0}
%\includegraphics[width=\textwidth]{LGB_c+n_cn.pdf} 
%\caption{[(C+N)/Fe] for the lower giant branch stars (2.6 $<$ log(g) $<$ 3.3). Points are color-coded by [C/N] abundance. Medians binned in [Fe/H] are plotted as black squares. A fiducial line from the Solar Neighborhood, $|$Z$|$ $>$ 0.5 kpc bin is plotted in all panels.}
%\label{fig:LGB_c+n}
%\end{sidewaysfigure*}

\begin{figure*}
\epsscale{1.0}
\plotone{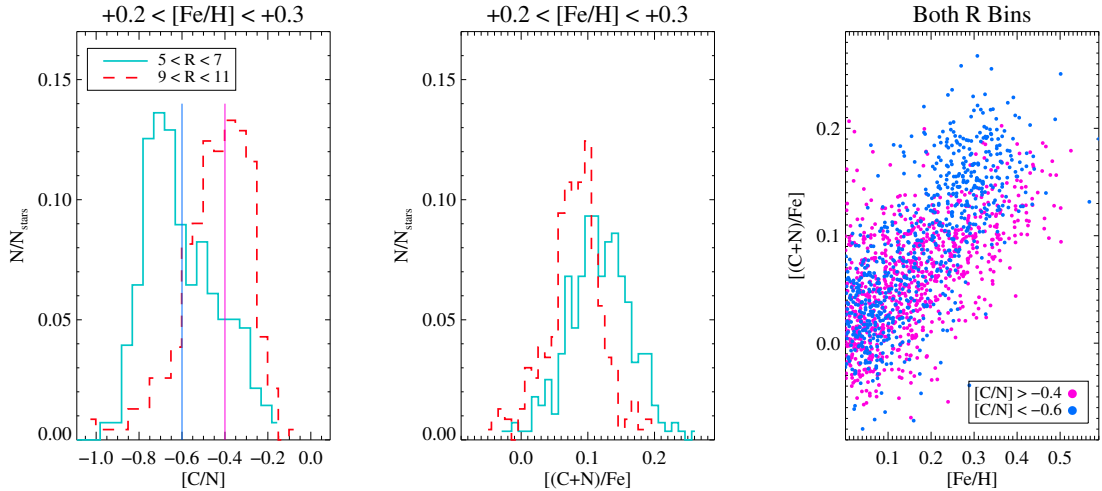} 
\caption{Elemental distributions for the LGB sample with $|$Z$|$ $<$ 0.5 kpc. Left: [C/N] distribution for the 5 kpc $<$ R $<$ 7 kpc bin (blue) and the 9 kpc $<$ R $<$ 11 kpc bin (red) for stars with +0.2 $<$ [Fe/H] $<$ +0.3. Middle: same as left but for [(C+N)/Fe]. Right: [(C+N)/Fe] vs. [Fe/H] for the two combined bins, but separated at [C/N] $>$ $-$0.4 (orange) and [C/N] $<$ $-$0.6 (cyan), indicated in the left plot by the vertical lines.}
\label{fig:LGB_cn_hist}
\end{figure*}

We take this one step further by analyzing the [C/N] and [(C+N)/Fe] distributions of the inner Galaxy and outer Galaxy for a small section in [Fe/H] (+0.2 $<$ [Fe/H] $<$ +0.3). The left panel of Figure \ref{fig:LGB_cn_hist} illustrates the different [C/N] distributions between the inner and outer Galaxy for the LGB sample, which we described as a sign of radial migration. The middle panel suggests that there may be a slight difference in the means of the two [(C+N)/Fe] distributions, however, when splitting the stars by [C/N] instead of radius, we see that both the low-[C/N] and high-[C/N] stars exhibit this trend of increasing [(C+N)/Fe] with increasing [Fe/H]. This again suggests that the [(C+N)/Fe] increase is mostly decoupled from the [C/N] abundance patterns.

If we interpret [C/N] as an age indicator, then both old and young stars show an increase in [(C+N)/Fe] with increasing [Fe/H]. This likely rules out both intermediate-mass and low-mass AGB ejecta as the source of this increase, as the timescales for enrichment are too long to contribute to the enrichment of the older stellar populations found in the inner Galaxy. Moreover, AGB yields from \citet{Karakas2016} actually suggest that the solar-metallicity models produce more C and N relative to Fe than the super-solar metallicity models, although more massive AGB stars produce more C and N relative to Fe than less-massive AGB stars. They also note that it is unlikely intermediate-mass AGB stars contributed significantly to the chemical evolution of the MW disk. \citet{Henry2018} calls this into question, noting that they find larger N/O abundance ratios in their observations of PNe than most models predict, but this enhancement does not appear to be a function of metallicity. However, \citet{vanZee1998} finds an increase of N/O with O/H in H II regions, which indicates that metal-rich Type II SNe begin to add nitrogen that would cause an increase in the total [(C+N)/Fe]. This latter scenario would explain the observed increase in [(C+N)/Fe] with increasing [Fe/H] observed in both the inner and outer Galaxy.

\subsection{Two-Infall or Major Merger}
\label{sec:infall}

Metal-rich old stars at the same Galactic radius as metal-poor young stars could, in principle, be a result of two major gas accretion events in the MW's history that were each followed by an epoch of star formation. The first epoch is a vigorous star-formation event spurred by an initial infall of gas that forms the high-$\alpha$ sequence of stars out to some radius. Then, there is a break in star formation followed by the accretion of gas that has been polluted in part by SNe from previous star-formation events (e.g., a ``Galactic Fountain'', \citealt{Marasco2013}), but also sufficiently diluted by pristine gas such that it is more metal-poor (but still solar-like $\alpha$-element abundance) than stars that had already formed in the MW. This results in younger stars being more metal-poor than the older metal-rich stars formed at the end of the first major star formation epoch that happen to reside at the same Galactic radius. A decrease of gas-phase metallicity with time can also arise in a chemical-evolution model with outflows, if the outflow efficiency increases with time (see, e.g., Fig.~9 of \citealt{Weinberg2017}).

Recent results from \citet{Grand2018} suggest that low-$\alpha$, low-[Fe/H] stars can form in the outer Galaxy if an initial epoch of star formation is followed by a paucity of star formation which results in a shrinkage of the gaseous disk to a smaller radius than the radius where it previously formed stars. Future accretion of pristine gas then mixes with Type Ia SNe remnant material which could, in principle, then form stars at a lower metallicity than the initial epoch of star formation. \citet{Mackereth2018} similarly find that such a star-formation history can give rise to the observed [$\alpha$/Fe] bimodality. However, they find that the median [Fe/H] of both the high-$\alpha$ and low-$\alpha$ sequence increases with time, implying that it is migration that causes the observed flattening of the age-metallicity relation in the MW.  Studies of the MW disk using the entire suite of APOGEE chemical abundances (which is beyond the scope of this paper) will shed more light on the two-infall scenario, and answer whether this scenario might naturally explain the observed [C/N]-[Fe/H] abundance trends, or if significant migration is indeed required. 

It is also possible that the old, metal-rich stars that are found throughout the Galactic disk are a result of a major merger of a relatively massive galaxy that was able to enrich its ISM to super-solar metallicities before merging with the MW. However, because these stars are as metal-rich as [Fe/H] $\sim$ +0.3-+0.4, such a galaxy would likely either have had to have been more massive than the LMC, or had an exotic star formation history such that it was able to enrich its gas to much higher metallicities than would be predicted by its mass based on the observed mass-metallicity relation of galaxies observed today (e.g., \citealt{Kirby2013}). The former scenario would then argue that the MW disk contains a large fraction (10+\%) of stars coming from a major merger. 

\subsection{Measurement Systematics}
\label{sec:meas}

In a recent comparison of the APOGEE abundances to optical abundances for an overlapping sample of stars, \citet{Jonsson2018} found that the difference between APOGEE [N/Fe] abundances and optical [N/Fe] abundances (from \citealt{daSilva2015} and \citealt{Brewer2016}) increases with increasing [Fe/H] such that the APOGEE [N/Fe] are $\sim$ 0.2-0.3 dex higher at [Fe/H] $>$ 0.3. While the APOGEE spectral region contains many more CN lines from which N abundances can be derived than the optical, it is not conclusive whether the APOGEE abundances or the optical abundances are systematically off. Such a systematic could be responsible for the increase of [(C+N)/Fe] with [Fe/H] seen in \S \ref{sec:AGB}, and observed by \citet{Masseron&Gilmore2015} in DR12. However, this potential measurement systematic would not explain the outer Galaxy stars being preferentially high in [C/N] but the inner Galaxy stars exhibiting a range in [C/N] at the same metallicity. 

\section{Conclusions}

We have presented APOGEE [C/N]-[Fe/H] abundance trends across much of the Galactic disk (3-15 kpc). Similar to other studies of APOGEE [C/N] abundances, we demonstrate that these abundance trends can be interpreted as age-[Fe/H] trends, allowing for the temporal exploration of the chemical evolution of the MW Galaxy.

The APOGEE LGB sample exhibits significant scatter in [C/N], which we interpret as scatter in age, across all metallicities in all regions of the Galactic disk. Therefore, we conclude that there is no age-metallicity relation in any part of the MW disk explored in this work. The [C/N] dispersion for stars with [Fe/H] $>$ $-$0.3 increases from 0.13 dex to 0.17 dex from out of the plane to in the plane. The increase is driven by the inclusion of the youngest stars (age $<$ 2.5 Gyr) that are only found in the plane. 

Far from the plane ($|$Z$|$ $>$ 1 kpc), we observe a gradient in the mean [C/N] of stars of $-$0.04 dex/kpc from 6-12 kpc, implying that the stars far from the plane become younger from inner to outer Galaxy. This is similar to the age gradient of the thick disk found by \citet{Martig2016b}, and consistent with the prediction of \citet{Minchev2015}, who show that such an age gradient in a thick disk is expected if the thick disk is a result of the flaring of mono-age populations. In the plane ($|$Z$|$ $<$ 0.5 kpc), there is a radial [Fe/H] gradient of the youngest stars of $-$0.060 dex/kpc, whereas the older stars exhibit a much flatter gradient of $-$0.016 dex/kpc. Both of these gradients are consistent with what \citet{Anders2017b} measured using the CoRoGEE sample. Such a flattening of the [Fe/H] gradient of the disk with time is predicted by \citet{Minchev2013} to be a consequence of radial heating and migration.

The UGB sample, which consists of stars with lower surface gravity that are few in number in the APOKASC catalog, appear to follow [C/N]-[Fe/H] abundance trends that are similar to those of the LGB sample. Notable differences are the smaller overall range in [C/N] as well as the [C/N]-[Fe/H] correlation for stars with [Fe/H] $<$ $-$0.6, which is not observed in the LGB sample. This latter observation is likely due to extra metallicity-dependent mixing that occurs as a star evolves further up the giant branch. We find that the metal-rich stars ([Fe/H] $>$ +0.2) that reside in the 5 kpc $<$ R $<$ 7 kpc bin are much younger, on average, than stars in the 3 kpc $<$ R $<$ 5 kpc bin. This suggests that the inner Galaxy is experiencing relatively little star formation as compared to the rest of the disk, or that the inner disk is dominated by old stars.

In the LGB and UGB sample, we also observe an upturn in the [C/N]-[Fe/H] abundance trends in the plane for R $>$ 7 kpc. This upturn is a consequence of the metal-rich stars ([Fe/H] $>$ +0.2) in the outer Galaxy exhibiting larger [C/N] values than more metal-poor stars in the same bin, suggesting they are older. Conversely, these metal-rich stars in the inner Galaxy exhibit a range of ages, but are skewed towards younger ages. It is plausible that radial migration is responsible for the presence of the metal-rich stars in the outer Galaxy that are $\sim$ 0.3-0.4 dex more metal-rich than the metallicity of the youngest stars in the outer Galaxy. Additionally, the [C/N] distributions of the Solar Neighborhood, which are peaked at intermediate ages across a range of metallicities (including the metallicity of the present-day Solar Neighborhood ISM), suggest that a majority of stars in the Solar Neighborhood formed farther in and migrated outward to the Solar Neighborhood. This is qualitatively consistent with the predicted distribution of birth and final radii from \citet{Minchev2014}. 

We have also explored other possible explanations for the observed [C/N]-[Fe/H] abundance trends. The large dispersion in [C/N]/age at fixed metallicity could be a consequence of a Galactic ISM that does not evolve substantially in metallicity over the past $\sim$ 4-6 Gyr, rather than radial migration. This could explain the large scatter in the ages of stars at solar metallicity in the Solar Neighborhood, but a mechanism like radial migration would still need to be invoked to explain the intermediate age stars that are 0.2-0.3 dex more metal-poor than the young metal-rich stars forming in the inner Galaxy, as well as the old metal-rich stars in the outer Galaxy. We also explore whether the [C/N]-[Fe/H] abundance pattern is a result of primordial variation by analyzing the [(C+N)/Fe] abundance, which is a quantity that should be conserved during stellar evolution, and is therefore indicative of a primordial [(C+N)/Fe] abundance. We do find that there is an increase of [(C+N)/Fe] with [Fe/H] at super-solar metallicities, but this increase occurs across the entire Galaxy and for all values of [C/N], suggesting that whatever is causing the increase in [(C+N)/Fe] does not significantly alter the [C/N] abundance. However, studies of C and N abundances of dwarf/subgiant stars across the entire Galaxy is necessary to properly quantify the primordial variation of [C/N].

\vspace{0.5cm}
\scriptsize{\emph{Acknowledgments.} 

We thank the anonymous referee for their insightful comments that really helped to improve this manuscript.

Funding for the Sloan Digital Sky Survey IV has been provided by the
Alfred P. Sloan Foundation, the U.S. Department of Energy Office of
Science, and the Participating Institutions. SDSS acknowledges
support and resources from the Center for High-Performance Computing at
the University of Utah. The SDSS web site is www.sdss.org.

SDSS is managed by the Astrophysical Research Consortium for the Participating Institutions of the SDSS Collaboration including the Brazilian Participation Group, the Carnegie Institution for Science, Carnegie Mellon University, the Chilean Participation Group, the French Participation Group, Harvard-Smithsonian Center for Astrophysics, Instituto de Astrof{\'{\i}}sica de Canarias, The Johns Hopkins University, Kavli Institute for the Physics and Mathematics of the Universe (IPMU) / University of Tokyo, Lawrence Berkeley National Laboratory, Leibniz Institut f{\"u}r Astrophysik Potsdam (AIP), Max-Planck-Institut f{\"u}r Astronomie (MPIA Heidelberg), Max-Planck-Institut f{\"u}r Astrophysik (MPA Garching), Max-Planck-Institut f{\"u}r Extraterrestrische Physik (MPE), National Astronomical Observatory of China, New Mexico State University, New York University, University of Notre Dame, Observat{\'o}rio Nacional / MCTI, The Ohio State University, Pennsylvania State University, Shanghai Astronomical Observatory, United Kingdom Participation Group, Universidad Nacional Aut{\'o}noma de M{\'e}xico, University of Arizona, University of Colorado Boulder, University of Oxford, University of Portsmouth, University of Utah, University of Virginia, University of Washington, University of Wisconsin, Vanderbilt University, and Yale University.

D. A. G. H. and O. Z. acknowledge support provided by the Spanish Ministry of Economy and Competitiveness (MINECO) under grant AYA-2017-88254-P. T.C.B. acknowledges partial support for this work from grant PHY 14-30152; Physics Frontier Center/JINA Center for the Evolution of the Elements (JINA-CEE), awarded by the US National Science Foundation.

% D.A.G.H. was funded by the Ram\'on y Cajal fellowship number RYC$-$2013$-$14182. D.A.G.H. acknowledge  support  provided  by  the  Spanish  Ministry  of  Economy  and  Competitiveness (MINECO) under grant AYA$-$2014$-$58082$-$P. T.C.B. acknowledges partial support for this work from the National Science Foundation under Grant No. PHY-1430152 (JINA Center for the Evolution of the Elements). D.M. is supported by the Basal CATA through grant PFB-06, and the Ministry for the Economy, Development, and Tourism, Program ICM through grant IC120009, awarded to the MAS, and by FONDECYT grant No. 1130196. SV gratefully acknowledges the support provided by Fondecyt reg. n. 1170518. S.R.M. acknowledges NSF awards AST-1109718, AST-1312863, and AST-1616636.

\normalsize

\bibliographystyle{apj}
\bibliography{ref_og.bib}
 
\end{document}